\documentclass[conference]{IEEEtran}

\ifCLASSINFOpdf
\else
\fi

\usepackage{xcolor}
\usepackage{graphicx}
\usepackage{amsmath}
\usepackage{mathrsfs}
\usepackage{float}
\usepackage{subcaption}

\usepackage{epsfig,endnotes}
\usepackage{xspace}
\usepackage{graphicx}
\usepackage{longtable}
\usepackage{booktabs}
\usepackage{enumitem}
\usepackage{amssymb}
\usepackage{soul}
\usepackage{ulem}
\usepackage{xurl}

\newcommand{\newpar}[1]{\vspace{1mm}\noindent\textbf{#1}\@\xspace}

\newcommand{\SP}{S\&P\xspace}

\begin{document}

\title{From Perception to Protection: A Developer-Centered Study of Security and Privacy Threats in Extended Reality (XR)}

\author{
    \IEEEauthorblockN{Kunlin Cai\IEEEauthorrefmark{2},
    Jinghuai Zhang\IEEEauthorrefmark{2},
    Ying Li\IEEEauthorrefmark{2},
    Zhiyuan Wang\IEEEauthorrefmark{5},
    Xun Chen\IEEEauthorrefmark{4}\thanks{A portion of this work was conducted during Kunlin Cai's internship with Xun Chen at Samsung Research America.},
    Tianshi Li\IEEEauthorrefmark{3}, and
    Yuan Tian\IEEEauthorrefmark{2}}
    \IEEEauthorblockA{\IEEEauthorrefmark{2}University of California, Los Angeles. Email: \{kunlin96, jinghuai1998, yinglee, yuant\}@g.ucla.edu}
    \IEEEauthorblockA{\IEEEauthorrefmark{3}Northeastern University. Email: tia.li@northeastern.edu}
    \IEEEauthorblockA{\IEEEauthorrefmark{4}Independent Researcher. Email: xunchen@outlook.com}
    \IEEEauthorblockA{\IEEEauthorrefmark{5}University of Virginia. Email: vmf9pr@virginia.edu}
}

\IEEEoverridecommandlockouts
\makeatletter\def\@IEEEpubidpullup{6.5\baselineskip}\makeatother
\IEEEpubid{\parbox{\columnwidth}{
		Network and Distributed System Security (NDSS) Symposium 2026\\
		23-27 February 2026, San Diego, CA, USA\\
		ISBN 979-8-9919276-8-0\\
	https://dx.doi.org/10.14722/ndss.2026.230807\\
		www.ndss-symposium.org
}
\hspace{\columnsep}\makebox[\columnwidth]{}}

\maketitle

\begin{abstract}
The immersive nature of XR introduces a fundamentally different set of security and privacy (\SP) challenges due to the unprecedented user interactions and data collection that traditional paradigms struggle to mitigate. As the primary architects of XR applications, developers play a critical role in addressing novel threats. However, to effectively support developers, we must first understand how they perceive and respond to different threats. Despite the growing importance of this issue, there is a lack of in-depth, threat-aware studies that examine XR \SP from the developers’ perspective. To fill this gap, we interviewed 23 professional XR developers with a focus on emerging threats in XR. Our study addresses two research questions aiming to uncover existing problems in XR development and identify actionable paths forward.

By examining developers' perceptions of \SP threats, we found that: (1) XR development decisions (e.g., rich sensor data collection, user-generated content interfaces) are closely tied to and can amplify \SP threats, yet developers are often unaware of these risks, resulting in cognitive biases in threat perception; and (2) limitations in existing mitigation methods, combined with insufficient strategic, technical, and communication support, undermine developers' motivation, awareness, and ability to effectively address these threats.
Based on these findings, we propose actionable and stakeholder-aware recommendations to improve XR \SP throughout the XR development process. This work represents the first effort to undertake a threat-aware, developer-centered study in the XR domain—an area where the immersive, data-rich nature of the XR technology introduces distinctive challenges.

\end{abstract}

\IEEEpeerreviewmaketitle

\section{Introduction}
With technological advancements, extended reality (XR) has become increasingly accessible in various aspects of daily life, offering immersive experiences that blur the boundaries between physical and digital worlds.
XR developers have been developing numerous XR applications which revolutionizing how people learn, work, and interact. 
By 2025, the global XR market is projected to reach USD 87.3 billion, with the applications segment accounting for the largest share at 65\%, signaling significant growth in XR applications~\cite{Shinde_2024}. 
However, the rapid expansion of the XR applications is a double-edged sword, offering substantial benefits while introducing notable security and privacy (\SP) risks~\cite{Soroushian_2021}.

The concerns related to \SP in XR applications---encompassing augmented reality (AR), virtual reality (VR), and mixed reality (MR)---require a dedicated investigation due to their unique characteristics:
(1) XR provides users with unparalleled \textit{immersive} experiences~\cite{Weinstein_2024} by enabling innovative interaction designs unique to spatial computing, such as lifelike avatar embodiment~\cite{smith2018communication} and gaze-based or emotion-driven interactions~\cite{plopski2022eye, speicher2019mixed}. These advancements introduced new security concerns, including but not limited to XR side-channel attack~\cite{yang2023can,su2024remote, zhang2023s}, immersive digital manipulation~\cite{cheng2023exploring, casey2019immersive, tseng2022dark}, identity threats~\cite{shi2021face}, and intellectual property threats from blending real and virtual content~\cite{giaretta2024security, de2019security}.
(2) XR inherently requires an extensive collection of \textit{multimodal user and environment data}, such as gestures, gaze, voice, physiological signals, location, and movement. The combination of these data streams provides a granular, real-time representation of users' state, where prior studies~\cite{nair2023truth,wierzbowski2022behavioural,nair2023unique,slocum2023going} have shown that such fine-grained data opens up new avenues for attackers to exploit XR systems and compromise user privacy.

The interaction designs and data collection channels in XR are primarily determined by developers at the application level. 
Since developers play a key role in creating these designs~\cite{acar2016you}, their design choices directly shape the \SP posture of applications. However, current XR research ~\cite{katins2024assessing,sb2025they} primarily focuses on user-centered studies, aiming to understand the \SP issues that users care about.
Although existing studies provide valuable insights from the user perspectives, they often fall short in connecting \SP issues with the actual development process and uncovering the key factors hindering \SP development.  
Therefore, there's an urgent need for dedicated developer-centered studies to gain a deeper understanding of the causes behind emerging threats in XR applications.

Specifically, developers' %
awareness, misconceptions, attitudes, and capabilities~\cite{acar2016you,horstmann2024those,gutfleisch2022does} in \SP can directly affect their development decision. However, only a few studies have examined \SP in XR development at a high level, often focusing on general concerns rather than specific threats. 
For example, Adams et al.~\cite{adams2018ethics} interviewed developers about their general concerns and attitudes about XR \SP.
These studies only rely on developers' subjective perceptions and concerns, which limits the identification and exploration of \textit{unknown unknowns}—that is, situations where developers are unaware of, or cannot systematically recall XR-specific threats (i.e., threats introduced by XR interaction design or data collection), and reflect on them within the context of their app development.
Given the emergence of diverse threats in XR, dedicated studies are needed to investigate these threats, examine how developers perceive them, and assess how these threats affect XR development practices.
The absence of a threat-focused, developer-centered study makes it challenging to understand: (1) The relationship between XR development and the presence of different threats in XR applications, (2) the challenges developers face in implementing effective mitigation strategies, and (3) the ways to provide actionable solutions to support developers in addressing these threats.

To address this research gap, our work takes an initial step toward providing an in-depth understanding of how developers perceive emerging threats and existing mitigation strategies in XR, aiming to fill the missing but necessary knowledge to handle both current and emerging threats.
More formally, we have two research questions:

\vspace{0.1cm}

\noindent\textbf{RQ1:} What are developers' perceptions of emerging \SP threats in XR? 

\vspace{0.1cm}
\noindent\textbf{RQ2:} What are the developers' perceptions of current mitigation practices and the support from the XR community?
\vspace{0.05cm}

\begin{figure}[t]
    \centering
    \begin{subfigure}{0.23\textwidth}
        \centering
        \includegraphics[width=\textwidth]{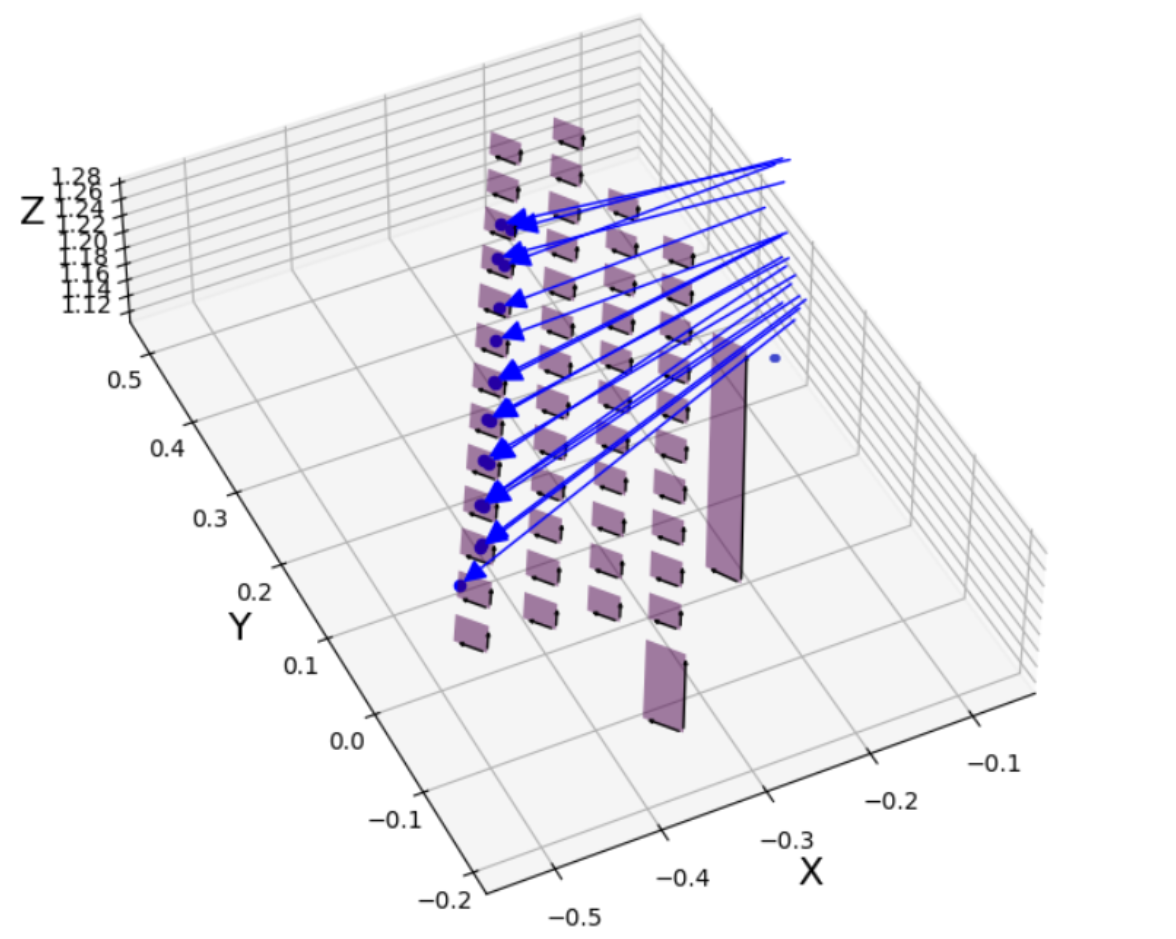}
        \caption{VR Keylogging Attack~\cite{su2024remote}}
    \end{subfigure}
    \hfill
    \begin{subfigure}{0.23\textwidth}
        \centering
        \includegraphics[width=\textwidth]{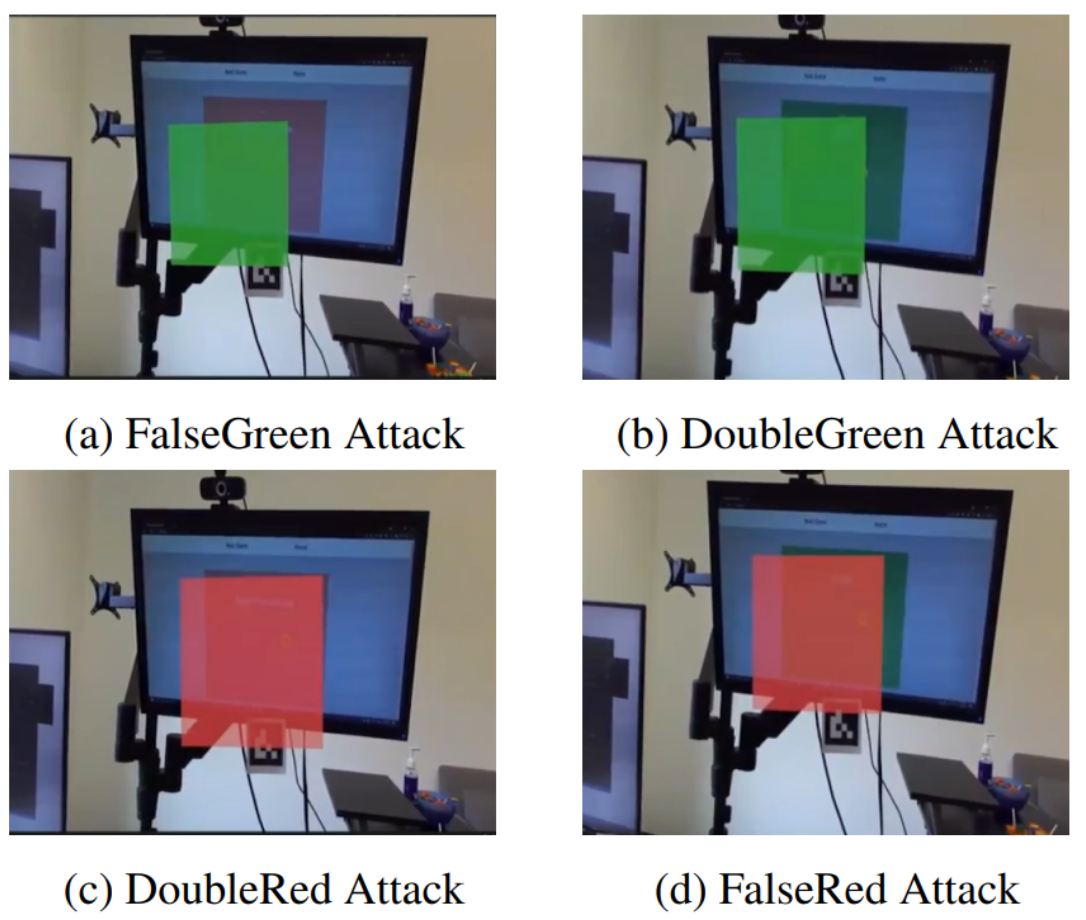}
        \caption{AR Perception Attack~\cite{cheng2023exploring}}
    \end{subfigure}
    \caption{Attack examples shown to participants (developers).}
    \label{fig:extraction_results}
    \vspace{-3mm}
\end{figure}

To address the research questions,
we conducted semi-structured interviews with 23 professional XR developers. %
To investigate developers’ perspectives on threats unique to or amplified by XR, we curated and presented a diverse set of sensitive data sources and attack scenarios (e.g., the attacks illustrated in Figure~\ref{fig:extraction_results}). These examples, sourced from top-tier conference publications, represent a broad and emerging threat landscape in XR and enabled us to elicit developer feedback.
We then asked developers to reflect on the necessary mitigations for these threats and the responsibilities various stakeholders in the XR community should undertake.
Following the interviews, we conducted a qualitative analysis using a bottom-up, open-coding approach to identify developers’ perceptions of different \SP threats and mitigation.

Our interviews reveal that developers recognize the importance of XR \SP but are hindered by awareness gaps and unmet supports from the XR community.
From developers' perspectives on the presented threats (Section~\ref{result_privacy},~\ref{result_security}), we identified factors that make it difficult for XR developers to recognize potential \SP threats. They include suboptimal practices caused by awareness gaps, sensitive use cases (e.g., the collection of rich sensor data from the user's viewpoint), and the tension between \SP and the immersive nature of XR environments (e.g., increased potential for user misuse). %
Furthermore, in Section~\ref{RQ2}, we found gaps in developers’ awareness of existing tools for threat mitigation, suggesting that developers may still fail to take effective actions to protect their users from known threats. We also observed that many community solutions either reduce utility (e.g., blocking raw camera feeds), rely on outdated documentation (e.g., Meta’s API docs), or require external \SP experts—impractical for XR developers, especially those at small companies with limited budgets. %
These external issues could exacerbate developers' challenges for mitigation in XR development, awareness gaps and suboptimal practices.

Our findings suggest that despite the growing number of \SP threats revealed in academic research and threats occur in real-world applications, XR developers face multiple challenges to keep up-to-date knowledge and promptly address them in their actual development process.
In Section~\ref{discussion}, we synthesize two core problems, \textit{awareness gaps} and \textit{diffusion of responsibility}, and discuss how the emerging nature of XR interactions and the user-experience-driven nature of XR threats contribute to these problems. Following these insights, we propose stakeholder-aware recommendations to enhance \SP in XR.

\noindent
We summarize our contributions as follows:~~ 
\begin{itemize}[leftmargin=*]
\item To the best of our knowledge, we are the first to conduct a developer-centered study exploring developers’ perceptions (e.g., awareness, misconceptions, attitudes, and capabilities) regarding diverse emerging threats in XR.
\item We conducted 90-minute interviews with 23 professional XR developers to identify contexts in which \SP threats are considered important or amplified during XR development (e.g., the amplification of sensitive data through immersive designs, the collection of rich sensor data, unmoderated XR social experiences, etc.).
\item We further identified gaps in developers' perceptions and knowledge about existing mitigations and support, and then proposed improvement directions from both developers’ and the broader community’s perspectives. These findings highlight the urgent need for enhanced support in XR \SP strategies, technologies, and communication.
\end{itemize}

\section{Background and Related Work}

\subsection{Security and Privacy in Extended Reality (XR)}
\label{background:xrsp}

XR encompasses emerging technologies that deliver immersive experiences, including Augmented Reality (AR), Virtual Reality (VR), and Mixed Reality (MR).
Recent research highlights that the unique characteristics of XR development make it particularly vulnerable to \SP threats~\cite{garrido2023sok}.
Specifically, XR applications collect an extensive range of fine-grained human data (e.g., motion, voice, and virtual behavior) which often contains more granular and sensitive user information than traditional applications.
Breaches in XR also pose more severe risks to user identity~\cite{nair2022going, nair2023unique, nair2023truth}, surrounding environments~\cite{farrukh2023locin, guzman2021unravelling}, and typing information~\cite{luo2024eavesdropping, su2024remote}. %
XR environments also introduce new attack surfaces that developers must consider when developing interactive designs. Researchers have identified XR-specific problems with overlooked design choices and specialized functionalities~\cite{su2024remote, yang2023can}, virtual social experience~\cite{blackwell2019harassment,freeman2022disturbing,jansen2020social,schulenberg2023creepy,zheng2023understanding}, content security~\cite{cheng2024user, slocum2024doesn,tariq2023deepfake}, user perception~\cite{cheng2023exploring,casey2019immersive,tseng2022dark}, side-channels~\cite{al2021vr, zhang2023s}, and insecure implementations~\cite{trimananda2022ovrseen, guo2024empirical}. 
Moreover, there are only a few standards, policies, and guidelines~\cite{cyberxr2021immersive,rosenberg23the} that address \SP in the context of XR development. Although these policies provide suggestions for avoiding basic security issues (e.g., identity, intellectual property) and privacy leakage (e.g., location, financial information, and personal preferences), they remain insufficient. They fail to cover many XR-specific threats~\cite{giaretta2024security} and types of sensitive data~\cite{garrido2023sok}, and they lack an understanding of the relationship between XR development and existing threats. Our work seeks to address this gap by providing a deeper understanding of XR developers, identifying \SP issues in XR development, and offering actionable directions for \SP-aware XR development.

\subsection{Developer-Centered Qualitative Study on \SP}
While user-centered studies offer valuable insights into the current situation, they lack an in-depth understanding of how development processes contribute to emerging \SP threats. To this end, research focused on developers’ perceptions—such as their awareness, misconceptions, attitudes, and capabilities—is essential, given developers' central role in the software lifecycle. 
Such developer-centered studies can better reveal the causes of \SP issues and inform the creation of usable, adoptable solutions.
For instance, Acar et al.~\cite{acar2016you} argue that researchers should consider developers in the \SP landscape. Building on this, researchers have examined how developers contribute to \SP issues, including their role in vulnerable implementations~\cite{votipka2020understanding, assal2018security}, opinions on privacy concerns~\cite{li2021developers, tahaei2021privacy}, and perceptions of \SP documents~\cite{horstmann2024those, li2022understanding}. 

However, there is a lack of developer-centered studies focusing specifically on XR \SP. A better understanding of how threats in XR \SP affect development is needed to identify existing problems and build stronger support around the development process~\cite{hine2024safety,10005232}. Such support is currently limited in XR, largely due to emerging threats associated with new interaction designs in XR.
Most recent XR \SP-related studies primarily focus on users~\cite{katins2024assessing, sb2025they,abhinaya2024enabling, cao2024understanding} and experts with notable experience and knowledge in XR~\cite{abraham2022implications}. While there is a general developer involved study~\cite{adams2018ethics} on VR developers' perspectives on \SP, it does not address XR-specific threats, such as those related to XR interactions and extensive data collection. 
Abhinaya et al.~\cite{abhinaya2024enabling} examined both users’ and developers’ perceptions of harassment in XR, uncovering several key concerns. However, by focusing exclusively on harassment, it does not compare or analyze other \SP threats, which limits its ability to surface broader development challenges, or offer actionable recommendations for emerging risks from the developers’ perspective.
A significant gap remains, as none of these studies examine developers’ perceptions or responses to emerging XR threats, and the challenges of designing \SP-conscious XR applications. As previously discussed, this gap needs to be addressed to better understand the causes of unique \SP issues in XR.
To this end, we summarized XR-specific threats from existing literature and integrated domain expertise to conduct the first developer-centered and threat-aware study, aiming to understand developers' perceptions, the reasons behind those perceptions, and to provide usable solutions.

\section{Methods}

\subsection{Recruitment}
We recruited 23 professional XR developers with experience in developing and distributing XR applications. %
Participants were recruited by searching on LinkedIn using the keywords of \textit{XR/VR/AR developers}. We assessed their qualifications by reviewing their LinkedIn profiles and personal CVs, and directly contacted those who met our criteria. 
Participants were required to have published at least one complete XR-specific project on a public distribution platform (e.g., Meta Quest) and to be currently employed as XR developers. %
Additionally, we confirmed their XR experience by asking about their past projects through an invitation email or message.
Once their experience were verified, we proceeded by inviting them for a Zoom interview. Of the 412 qualified developers we contacted (from June 2024 to June 2025), 34 participants signed up, and 23 ultimately participated in the study. The decision to finalize the number at 23 was based on reaching saturation of themes, where no new themes emerged from additional interviews. All 23 participants completed the main study (approximately 90 minutes) and were compensated \$70 USD.

\newpar{Participant Demographics:}
Among our participants, eight identified as female, fourteen as male, and one preferred not to disclose their gender. The group was relatively young, with eight participants aged 18-25, eleven aged 25-35, three aged 35-45, and one aged 45-55. Despite randomly inviting all qualified participants we found online, most were young males, reflecting the gender and age makeup of the XR developer community, as observed in similar studies~\cite{karre2019virtual,adams2018ethics}.
Table~\ref{tab:demographic} shows the information on our diverse participant pool.
Specifically, our participants have experience developing 15 types of XR applications across different domains. %
Moreover, our participants have published applications on 12 different platforms.
Among all of our participants, only two participants have less than 1 year of experience, six have 1-2 years of experience, eight have 2-5 years of experience, six have 5-10 years of experience, and one has more than 10 years of experience.
Notably, two participants have contributed to XR applications with over one million downloads.

\begin{table*}[t!]
\renewcommand{\arraystretch}{1.2}
\fontsize{7.5}{7.5}\selectfont
\centering
\caption{\textbf{Participant Overview} -- Years of experience in XR development, number of XR applications developed, the categories and styles of developed applications, and the platforms where these applications were published and accessed.}
\label{tab:demographic} 
\begin{tabular}{ccccccc}
\toprule
ID & XR Development YOE & \# of Apps & App Categories & App Style &  Publish Channel \\
\midrule
P1 & 0-1 Years & 1 & Single player & Education & Meta\\
P2 & 2-5 Years & 2 & Single player & Design, Action, Puzzle, Fitness &  Meta\\
P3 & 1-2 Years & 4 & Single \& Multi-player & Design, Education &  itch.io\\
P4 & 2-5 Years & 12 & Single \& Multi-player & Social, Action, Role-playing & Apple, Meta \\
P5 & 1-2 Years & 4 & Single \& Multi-player & Design, Action, Role-playing, Puzzle & Steam, itch.io \\
P6 & 1-2 Years & 3 & Single player & Retail, Sport & Snap \\
P7 & 5-10 Years & 2 & Single \& Multi-player & Social, Music & Steam, Meta \\
P8 & 5-10 Years & 5 & Single \& Multi-player & Social, Event & WebXR \\
P9 &5-10 Years & 7 & Single \& Multi-player & Education, Healthcare, Action, Sport & Meta, Steam, Pico, WebXR \\
P10 & $>$10 Years & 15 & Single \& Multi-player & Social, Education, Retail, Design, Action, Sport & Meta, Apple \\
p11 & 2-5 Years & 4 & Single \& Multi-player & Design, Action, Puzzle & Spark AR \\
P12 & 2-5 Years & 5 & Single player & Design, Action, Puzzle  & Meta, Apple, Google Play \\
P13 & 5-10 Years & $>$40 & Single \& Multi-player & Social, Education, Healthcare, Retail  & Meta, Directly to business \\
P14 & 1-2 Years & 3 & Single \& Multi-player & Social, Design, Action & Meta\\
P15 & 2-5 Years & 8 & Single \& Multi-player & Education, Healthcare, Retail, Action & Steam, itch.io\\
P16 & 2-5 Years & 2 & Single player & Education, Healthcare, Relax & Meta, Steam, Pico, itch.io\\
P17 & 0-1 Years & 8 & Single \&Multi-player & Social, Museum, Other & itch.io\\
P18 & 5-10 Years & 29 & Single \& Multi-player & Social, Education, Healthcare, Design, Puzzle & Steam, Apple, WebXR\\
P19 & 5-10 Years & 2 & Single player & Education, Design, Social & Meta, Apple \\
P20 & 1-2 Years & 3 & Single player & Retail, Design & Meta, Apple\\
P21 & 2-5 Years & 6 & Single player  & Education, Social AR Filter, Other & Snap, YouTube, Instagram, Meta\\
P22 & 1-2 Years & 3 & Single \& Multi-player & Education, Social & WebXR\\
P23 & 2-5 Years & 20 & Single player & Education, Design, Action & Meta\\

\bottomrule
\end{tabular}
\end{table*}

\subsection{Study Design}
\label{studydesign}
After recruitment, the study included a pre-study survey and an interview phase. The pre-study survey collected participants’ demographic information, development experience, and application publishing channels. %
The survey was completed within three days prior to each scheduled interview.

During the main study session, we conducted semi-structured interviews with each participant to gather information on their development practices, perceptions of the threats we demonstrated, and opinions on the XR industry's security and privacy (\SP). We assured participants of anonymity and emphasized that our study aimed to address issues affecting the entire developer community, not to test individuals. %

This study was approved by the Institutional Review Board. At the beginning of each study session, the interviewer briefed the participants on the study goals and procedures and then asked them to sign a consent form.
The semi-structured interview contains the following four modules:%

\newpar{Background Questions:} %
The interviewer asked participants about their XR development background, including training, involvement in \SP design, experiences with \SP issues, and perspectives on community practices. To elicit unique insights, the interviewer tailored prompts using participants’ pre-study survey responses (e.g., application details). Both prepared and impromptu follow-up questions were used to explore topics such as their role in \SP-related decisions. %

\newpar{Threats in XR:}
The second part of the interview aimed to understand participants' perceptions of \SP-related threats in XR, including sensitive data, which we extended based on~\cite{garrido2023sok} (Table~\ref{table:unique_sensitive_data_xr}), data leakage channels (Table~\ref{tab:channel}), and potential attacks.  
As detailed in Table~\ref{table:attacks_in_xr}, we reviewed notable and representative attacks documented in existing XR \SP literature from top venues, classifying them as XR-related (e.g., perception manipulation) or XR-amplified (e.g., social attacks). 
Building on the definitions proposed in~\cite{DARPA_2023b} and~\cite{giaretta2024security}, we organized these threats into categories for clarity and demonstration convenience. Initially, we categorized them based on primary attack purposes outlined in~\cite{DARPA_2023b}, then refined the categorization by incorporating variations in attack surfaces related to XR design choices and use cases.

We then presented these threats (e.g., perception, social attacks in XR) to participants in two different orders (i.e., ascending and descending) to gather their feedback. Due to time constraints, the demonstration included all threat categories but did not exhaustively cover every XR \SP threat identified in the literature. However, as our aim was to explore developers’ perceptions, we believe the selected examples were sufficient and broadly representative for XR threats.
After the demonstration, the interviewer asked participants to reflect on the threats regarding their practicality and importance. The interviewer followed up with questions to assess their comprehension of the content and ask for potential mitigation.

\begin{table*}[ht]
\fontsize{7.5}{7.5}\selectfont
\centering
\caption{Sensitive data in XR we collected and summarized from the literature~\cite{garrido2023sok,Pahi_2023}. }\label{table:unique_sensitive_data_xr}
\begin{tabular}{cll}
    \toprule
    \textbf{No.} & \textbf{Data Type} & \textbf{Sensitive Information Examples} \\ 
    \midrule
    1 & Motion-related data (e.g., motion tracking/gesture)~\cite{nair2023truth,nair2023unique, wu2023privacy} & Identity, age, gender, physical state, biometric information, etc. \\ \midrule
    2 & Voice data (e.g., microphone data)~\cite{luo2024eavesdropping, farrukh2023locin, cayir2025speak} & Speech content, gender, biometric, etc. \\ \midrule
    3 & Camera data (e.g., surrounding environment)~\cite{garrido2023sok,chandio2024stealthy} & Environment, personal information, identity, other person identity, etc.\\ \midrule
    4 & Device related data (e.g., network/HMD)~\cite{al2021vr, garrido2023sok, trimananda2022ovrseen} & Wealth, network, IP address, etc.\\ \midrule
    5 & Spatial data (e.g., location/play area)~\cite{garrido2023sok} & Wealth, identity, address, etc. \\ \midrule
    6 & Health related data (e.g., heart rate) ~\cite{Pahi_2023,zhang2023facereader,ye2025bpsniff} & Cognition, emotion, health, etc. \\ \midrule
    7 & Biometric (e.g., eye tracking/gait)~\cite{noah2024privacy,jarin2023behavr,wang2024gazeploit} & Attention, identity, authentication related information, etc.\\ \midrule
    8 & Observation (e.g., avatar/behavior)~\cite{su2024remote,yang2023can} & Age, gender, preference, health, emotion, etc. \\ \midrule
    9 & Feedback (e.g., haptics)~\cite{gugenheimer2022novel,Pahi_2023} & Virtual behavior, users state, etc. \\
    \bottomrule
    \label{data}
\end{tabular}
\end{table*}

\setlength{\tabcolsep}{2pt}
\begin{table*}[ht]
\fontsize{7.5}{7.5}\selectfont
\centering
\caption{Security and privacy attacks in XR we summarized from the literature.}\label{table:attacks_in_xr}
\begin{tabular}{clll}
    \toprule
    \textbf{No.} & \textbf{Attack Category} &\textbf{Attack Targets}& \textbf{Explanation} \\ \midrule
    1 & Shoulder Surfing Attacks &Spy XR user~\cite{gopal2023hidden}, Spy on bystanders~\cite{denning2014situ}, etc. & Recording without other people's consent \\ \midrule
    2 & Software Side-channel Attacks& Infer keystroke~\cite{su2024remote,yang2023can}, re-identify user~\cite{nair2023unique}, etc.& Exploiting functionality data as side-channel \\ \midrule
    3 & Input Attacks & Compromise input integrity~\cite{lebeck2016safely, dastgerdy2024virtual,giaretta2024security,shang2020secure}, DoS~\cite{cayir2023augmenting,el2024cybersecurity}, etc.& Manipulating inputs to trigger dangerous operation \\ \midrule
    4 & Social Attacks & Harass another user~\cite{blackwell2019harassment,freeman2022disturbing,schulenberg2023creepy,zheng2023understanding}, social engineering~\cite{jansen2020social}, etc.& Utilizing virtual social experience as attack channels \\ \midrule
    5 & Content Attacks& Overlay malicious content~\cite{cheng2024user, slocum2024doesn}, impersonate others~\cite{tariq2023deepfake}, etc.& Introducing malicious virtual content in XR\\ \midrule
    6 & Perception Attacks & Manipulate perception~\cite{casey2019immersive, cheng2023exploring, tseng2022dark}, manipulate memory~\cite{bonnail2023memory} etc.& Manipulating perception and memory \\ \midrule
    7 & Physiology Attacks& Introduce disorientation~\cite{casey2019immersive}, introduce content dizziness~\cite{freeman2018automated}, etc.& Introducing motion sickness and discomfort \\ 
    \bottomrule
\end{tabular}
\end{table*}

\newpar{Mitigation and Best Practices:}
The third part of the interview focused on understanding participants' awareness of existing mitigation strategies in XR and their perspectives on the responsibilities of various stakeholders (e.g., policymakers, users, platform providers) in the XR community. We compiled tools and practices from existing literature into a slide deck as provided in Table~\ref{table:approaches_xr_security_privacy}, asked developers about their awareness of these mitigations, and presented explanations and examples. We then asked participants to use the tools to address the threats mentioned previously and assess their understanding.

\subsection{Qualitative Analysis}
\label{method:label}
In alignment with our two research questions, we conducted a qualitative analysis of interview transcripts and screen recordings using a bottom-up open coding approach facilitated by MAXQDA. 
Following the best practices by Saldana~\cite{saldana2021coding}, our analysis involved two rounds of coding to ensure a thorough examination and identification of key themes.

Three researchers independently analyzed four interviews in the first round of coding to develop an initial codebook through repeated transcript readings and iterative discussions.
Daily meetings were held to confirm, refine, and merge codes, resulting in a preliminary codebook comprising 124 codes.
Building on this foundation, the researchers engaged in axial coding analysis. 
This phase involved merging similar codes and organizing them into overarching themes, providing a structured framework to address the research questions.

Following the initial coding process, the task of coding the remaining data was undertaken by three researchers using the agreed-upon codebook.
Each data sample was then coded by a primary coder and verified by another researcher.
Any new codes emerging from this coding process were discussed among all three researchers and incorporated into the codebook upon reaching consensus.
Following the recommendations of McDonald et al.~\cite{mcdonald2019reliability}, inter-rater reliability was not calculated for the coding process as the primary objective was to identify emergent themes rather than to seek coder agreement. 
This approach allowed the researchers to remain open to novel insights and patterns within the data.

The final codebook consists of 81 codes grouped into 18 themes. The complete codebook is provided in Appendix~\ref{appendix:codes}.

\label{criteria}
\newpar{Labeling Criteria:}
During our study, we analyzed the ratings of user responses to understand the relationship between developers' awareness and their threat perceptions. The primary labeling was performed by an author with expertise in XR \SP. To mitigate potential bias and ensure consistency, each response was independently coded by a second researcher, selected from a pool of three co-authors on the paper. The second coder then discussed the labeling decisions with the XR expert to reach an agreement. This approach reduced the risk of prejudice toward individual participants and improved the reliability of the labeling process. 
The interrater reliability scores are as follows: (1) awareness of attacks, we obtained a Cohen’s Kappa of 0.812; (2) for awareness of mitigation, the Cohen’s Kappa is 0.920; (3) and for mitigation quality, the quadratic weighted Kappa is 0.827. These scores indicate a high level of consistency among coders~\cite{mchugh2012interrater}.
\begin{itemize}[leftmargin=*,topsep=1pt, partopsep=2pt, itemsep=1pt]%
\item Awareness of attacks: We classified participants as ``Aware'' of an attack if they mentioned it without our prompt, stated that they had considered such an attack before, or provided additional examples of similar attacks. 
\item Awareness of mitigations: Participants were marked ``Unaware'' of a defense mechanism if they mentioned being unaware of it or misused it for solving demonstrated attacks. 
\item Quality of developer-proposed mitigation: %
The researcher rated the mitigation strategies proposed by developers on a \textit{scale of three}, where \textit{one} indicates strategies that do not make much sense, \textit{two} indicates strategies that make some sense but are missing important details, and \textit{three} indicates strategies that are well-thought-out or align with realistic solutions adopted in the industry. 
\end{itemize}

\noindent\newpar{Threat Model:}
Our threat model focuses on (1) external adversaries (e.g., network/system intruders, malicious XR users, bystanders, or third‑party API providers capable of generating harmful content) and (2) careless developers who may inadvertently introduce attack vectors or harmful content to XR applications. We do not consider malicious developers as we assume study participants do not have malicious intent.

During the presentation of each attack, we further discuss the specific attack channels and attacker capabilities. For example:
(1) Many XR platforms and applications (e.g., VRChat) allow user-generated content (e.g., custom rooms or game objects), which—if not properly validated—can enable content, perception, or physiology attacks.
(2) Malicious third-party API providers can introduce input, content, or software side-channel attacks when their services are integrated without proper vetting.
(3) External adversaries may exploit hardware vulnerabilities to manipulate sensor inputs or engage in malicious activities within social XR applications, leading to shoulder-surfing, input-based, or social attacks.

\section{Developers' Perceptions on Privacy in XR}
\label{result_privacy}
XR technologies typically involve a wide range of user data, which introduces novel privacy threats. This section explores developers' perceptions of these threats, particularly those arising from the unique characteristics of XR user data. In our pre-study survey, developers rated privacy in XR as highly critical, with a median Likert score of 6/7; only three developers rated it 4 (moderately critical) or lower.
With this in mind, we further investigate our research question on developers' perceptions of privacy in XR from three perspectives, following the demonstration described in Section~\ref{studydesign}:
\begin{itemize}[leftmargin=*, topsep=1pt, partopsep=2pt, itemsep=1pt]
    \item Developers' awareness of sensitive data in XR
    \item Developers’ perceptions of sensitive data in XR on: \begin{itemize}
        \item What makes data overall more sensitive?
        \item What makes certain data perceived as less sensitive?
        \end{itemize}
    \item Developers' perceptions of leakage channels in XR on: \begin{itemize}
        \item What makes leakage channels overall more realistic?
        \item What makes leakage channels perceived as less realistic?
        \end{itemize}
\end{itemize}

Additionally, we also analyze potential developer misconceptions at the end of this section.
Understanding these perceptions is essential, as their awareness, misunderstanding, and capability directly influence their practices, application designs, and protection mechanisms in XR~\cite{acar2016you,horstmann2024those}.

\subsection{Developers have limited awareness of sensitive data in XR}
\label{awareness:privacy}
When asked about the sensitive data types used or collected in their applications—--prior to being introduced to the definitions of sensitive data types and sensitive information in XR (Table~\ref{data})—--developers, on average, identified only \textit{2.1} types of XR-sensitive data type (e.g., motion, biometric) and \textit{0.5} types of non-XR data type (e.g., email, password) as being collected in their applications. 
This initial finding raises concerns about whether developers' applications indeed collect only this limited set of data types or if developers fail to recognize certain types of data as sensitive.

\newpar{Developers come to realize that their applications collect more types of sensitive data than they had previously thought.} 
After being presented with XR-sensitive data definition, developers identified \uline{significantly more XR-sensitive data types}, recognizing an average of \textit{4.8} types collected in their applications and acknowledging protection needs.
P3 reflected: \textit{``It makes you realize that some data that seems pretty harmless is actually a lot more potent,''} highlighting potential risks from developers' lack of awareness.

\subsection{What makes data overall more sensitive in XR}
\label{results_data_sensitive}
After being presented with various types of sensitive data (Table~\ref{data}) and their implications, developers were asked to rate the sensitivity of each data type. 
From their ratings, we found a median Likert score of 5.0/7.0. 
We analyzed their reasoning and identified the key factors as follows. %

\newpar{Immersive interaction reduces the awareness of data leakage.}
P4, P12 and P23 mentioned that the \uline{immersive interaction design} in XR makes voice data more prone to leakage and increases its sensitivity: \textit{``When you talk to people in VR, it brings back the same feeling that you're talking to people in real life, so often people don't have the expectation that they are being recorded.''}
Additionally, P7 and P8 raised concerns about unintentional self-disclosure in XR applications due to non-intuitive interaction design. Informed by their prior experiences in XR, P8 noted that this could increase the sensitivity of voice and expressed an intention to incorporate clearer notification mechanisms in his future applications as mitigation: \textit{``I did not realize that they could hear me. There was no point at which I was told or made aware of, or had visual feedback on the fact that I had an open microphone.''}

\newpar{Advanced tracking sensors produce sensitive data with greater details.}
P7, P8, P13-16, and P19-22 highlighted that data collected in XR contain \uline{rich sensitive information}, which allows adversaries to infer user information. P13, a healthcare XR developer, emphasized that XR technology amplifies the sensitivity of data leakage in healthcare applications: \textit{``Because this [XR] device involves a lot of activity and emotion expressing capability. Using the controller like with the phone, you just use the vibration or some haptic features to identify things, but XR can track my hand motion, and in future it can comes with some sensor for your legs, or even if not, the sensors available in the excess can still detect your leg movement and the energy and the pace you move.''}
Similarly, P13 highlighted that XR camera data is more sensitive than data from other devices as it provides more \uline{detailed information} and is physically attached to the users' direct view. %
Considering the sensitivity of the camera data, P8 mentioned that they would only collect this data for valid reasons and properly protect (e.g., via encryption) in development.

\newpar{The value of XR data incentivizes extensive collection and surveillance.}
P7, P13, P14, and P19-21 further noted that the \uline{economic value} of the sensitive information inferred from the data types collected in XR can drive XR headsets, applications, and service providers to collect user data (e.g., camera and spatial data.), significantly increasing the privacy sensitivity of these data.
P13 noted that scanning a room for guardian system with an XR headset can leak data to the platform:\textit{``Where is my room? Where is my couch? Where is my table? Where is my TV? So this data definitely has a business value. Because, let's say most of the ADs now understand what we talk and what I'm interested in, and based on that, they will publish those ADs, right? ... at the same time, certain people are not comfortable with sharing that information.''} 
These concerns align with findings in~\cite{trimananda2022ovrseen, gdpr_advisor_ar_advertising}, which show that data is already being collected for profiling, analytics, and personal advertisements in XR.
More critically, P7, P8, P16, and P21 expressed concerns that headset providers or governments could exploit camera and biometric data for surveillance purposes since these data provide detailed information about XR users and their private space (e.g., home).

\newpar{Opaque XR infrastructures impede effective mitigation.}
P4, P9, and P13 highlighted that XR devices are increasingly being used for medical purposes, but the \uline{black-box XR systems} present challenges related to data protection and compliance, thereby increasing the sensitivity of medical data. 
P9 mentioned the challenges he faced when shipping his applications with devices as a medical service, particularly in meeting compliance requirements from government agencies (e.g., VA) which require full visualization on collected data in devices~\cite{va_publications}: \textit{``The problems we have with the headsets are that it is bytedance, or it's Meta, HTC. There's no open source. But on the same day, you know, it's like we're following the API callback to mainland China, ... We're part of the VA, the US government doesn't want that.''}
\subsection{What makes  certain data less sensitive}
\label{data_not_sensitive}
While developers generally perceive data as more sensitive in XR, they still rate some data types as less sensitive (score $<$ 4). we analyzed their reasoning and identified the key factors: %

\newpar{The perceived benefits tend to overshadow the risks of privacy leakage.}
P4, P7, and P17 described a trade-off, classifying functionally essential XR data types as less sensitive and prioritizing \uline{technical performance over privacy concerns}.
P4 stated: \textit{``I think it's sensitive, but it's required for the application to work. It's hard to make a gesture-based game if you can't get the gesture.''} 
Similarly, the importance of motion data for enabling immersive experiences contributes to P7's perception of it as less sensitive: \textit {``If you want to exist in the same place as somebody else. You're gonna need to know what their hands are doing with their bodies, and people voluntarily add extra trackers to their bodies to have the avatar be more realistic.''}
This contradicts the finding in~\cite{nair2023truth}, where the connection between motion data and crucial XR operations increases the potential to infer important information.

\newpar{The insufficient knowledge of developers tends to obscure the consequences of privacy leakage.}
P1, P2, P10, P13, P15, P16, P18, P20, and P21 stated that they do not consider certain data types (e.g., motion, feedback) to be highly sensitive even after our demonstration, as they could not identify plausible risks of sensitive information leakage based on their experience. %
For example, P7 mentioned: \textit{``I don't think it's super sensitive. I think most kinds of data that you're gonna get from that, it's like, Oh, this person whacked their controller into a wall. And so there's a wall there.''} 
Moreover, P4, P10, P13, and P16 admitted that they were \uline{not aware} and had not previously considered the potential consequences of using or collecting feedback data (e.g., haptics). Therefore, they perceive feedback data as less sensitive as P13 said: \textit{``So this is a very niche and new thing, which we don't know, because the applicant is completely new, and people are still figuring out how to use. It's hard for me to say how much threat from my experience.''}

\subsection{What makes leakage channels more realistic in XR}
\label{leakage_channel_realistic}

After presenting potential sensitive data types and possible data leakage channels, we collected developer feedback on which channels are more likely in XR. We identified high realism ratings with a median Likert score of 6.0/7.0. Analysis of
their reasoning identified the key factors as follows:

\newpar{Prevalence of development misoperations due to technological immaturity.}
P4, P6, P7, P9, P10, P12, P13, P16, and P18-22 reported, based on their own experiences or those of XR developers they worked with, developers tend to \uline{blindly rely on third-party packages} (e.g., Photon~\cite{exitgames_photon}) when developing XR applications. For example, %
P12 noted: \textit{``It's a fact that people use APIs all the time without really knowing what's under it ... If you're using Unity to build anything, you're gonna use a bunch of packages.''}
P7, P9, P10, P12, P13, P15, and P16 also noted that insecure operations, as a data leakage channel, are likely to occur in XR.
P13 believes the \uline{immaturity of XR} technology and market makes this trend especially concerning: \textit {``When it comes to the AR/VR set of things. It's still a maturing technology where a lot of freelancers, small agencies, people from all sizes of teams, and backgrounds are coming in.
So, as I mentioned, no matter what the protocols we bring in, at the end of the day. It's a responsibility of the developing team.''} 
These threats stem from the current landscape of the XR industry, which is largely composed of individual developers and small enterprises~\cite{xr4europe2025}. This context suggests that developers should be careful when using third-party packages to avoid potential data breach.

\newpar{Ease of self-disclosure by users in an immersive environment}
P1, P4, P5, P7, P9, P13, P14, P17 and P21 expressed concerns that 
beyond intentional information sharing, XR’s immersive nature can more easily lead to the unintentional disclosure of private information (e.g., behavior patterns, voice).
For example, P17 noted that users may accidentally leak private data and emphasized that developers should take responsibility for helping to protect them: \textit{``I am a firm believer that users are stupid. So if you, if you have something, chances are they will screw it up. Doesn't mean it's intended. But there's always something right?''}
P14 emphasized that such misuse is more likely to occur in XR due to its immersive interaction, which tends to \uline{encourage self-disclosure}, aligning with the findings in~\cite{sykownik2022something,maloney2020anonymity}: \textit {``In VRChat. I think a lot of times we are interacting with friends in a very natural way. I wouldn't pay extra attention to how I behave. So that data could be very suggestive of who I really am.''} %

\subsection{What makes certain leakage channels less realistic in XR}
\label{channel_not_realisitc}
While developers generally agree with the presented leakage channels as realistic, we identified cases where they give a lower score (score $<$ 4). To better understand the reasoning behind these ratings, we classified developers' perception of leakage channels being less realistic due to:

\newpar{The responsibility is not acknowledged in XR development.}
P3, P6, P7, P9, P10, P13, P15, P16, and P21-23 mentioned that they were \uline{unfamiliar with} hardware-side channels for data leakage. For example, P3 stated: \textit{``I am not too knowledgeable about this one, so I'm giving a lower score.''}
Additionally, P16 believed that developers should not be responsible for it: \textit{``I think that is possible. But I'm not really concerned about that. And also as a developer, I can't really do much about that.''} Similarly, P7, P15 and P17 suggested that this issue should be addressed by XR headset providers and governments. %

\newpar{The belief that local XR applications are inherently secure.}
When asked about the necessity of implementing protections (e.g., encryption or secure data storage) to prevent XR data leakage in their applications for the types of data demonstrated in the study, P1-3, P6, P7, P10-12, P16 and P17 believed that data handled by local XR applications (i.e., those without network functionality or limited to single-player use) would be inherently secure, and therefore would not require additional security measures.
For example, P11 commented:  \textit{``I think it's just fine because it's a local APP. So it only processes those data and uses it, but doesn't actually send it to anything.''}

\subsection{Analysis: Potential misconceptions on privacy in XR}

Based on the developers' perceptions above, we analyze and summarize the following misunderstandings about XR privacy:

\noindent\newpar{Misconceptions regarding data sensitivity.}
From Section~\ref{data_not_sensitive}, we observed two common misconceptions: (1) the belief that the data with higher utility is less sensitive, and (2) the assumption that unfamiliar data is also less sensitive.
However, according to the GDPR~\cite{gdpr} and widely accepted research guidelines~\cite{lehigh_data_security_privacy, ucf_research_data_privacy, leiden_personal_sensitive_data}, any data that can be used to reveal an individual’s identity or personal information (e.g., financial or health data) should be classified as sensitive and protected using appropriate mechanisms (e.g.,~\cite{nair2022going}).

\noindent\newpar{Misconception on the data leakage channels.}
FromSection~\ref{channel_not_realisitc}, we observed the following misconceptions:
(1) Misconceptions of assuming that mitigating data-leakage channels (e.g., hardware side channels) is outside their scope. For example, developers may believe this responsibility lies with hardware teams or other stakeholders. However, prior work emphasizes that software developers should play an active role in preventing leakage, even when it originates in hardware. Developers are encouraged to participate in co-design efforts and implement software-level protections such as isolation~\cite{dangwal2020sok, yang2024hardware, dubey2023hardware}.
(2) The misconception that local XR applications are inherently secure. While local applications may offer more security than online apps, they remain vulnerable to leakages. For example, through unsafe Android API calls, side channels (e.g., rendering and motion), and insecure local storage~\cite{guo2024empirical}.

\section{Developers' Perceptions on Security in XR}
\label{result_security}
In addition to privacy threats, %
In this section, we explore developers’ perceptions and understanding of threats happening in XR.
Our pre-study survey reveals that developers generally perceive security in XR as a critical issue, with a median Likert score of 5/7. Among the 23 developers surveyed, only four rated XR security $<=4$ (Moderately Critical).
Building on these findings, we further investigate our research question on developers' perceptions of security in XR from three aspects, following the demonstration in Section~\ref{studydesign}:
\begin{itemize}[leftmargin=*]
    \item Developers' awareness of security attacks in XR
    \item Developers’ perceptions of attack importance regarding: \begin{itemize}
        \item What makes attacks in XR overall more important?
        \item What makes certain attacks perceived as less important?
        \end{itemize}
    \item Developers’ perceptions of attack practicability: \begin{itemize}
        \item What makes attacks in XR overall more practical?
        \item What makes certain attacks perceived as less practical?
        \end{itemize}
\end{itemize}
Similar to the previous section, we analyze these responses and highlight potential misconceptions held by developers.
We focus on these questions because developers' awareness, misconceptions, and capabilities directly influence the security and mitigation they adopt in XR applications~\cite{acar2016you,horstmann2024those}. %

\subsection{Developers exhibited limited awareness of XR attacks.}
\label{awareness:security}
Without our demonstration of the categorized XR attacks, developers recalled an average of only \textit{0.9/7.0} XR-specific attack types listed in Table~\ref{table:attacks_in_xr}. Furthermore, 6/23 developers reported being unaware of any XR-specific attacks. These findings suggest that, without reminders, developers may struggle to adequately consider potential XR-specific threats during the design and development of their applications. To better understand this gap, we further investigate whether the lack of awareness stems from developers never having encountered these threats or simply needing a prompt to recall them.

\newpar{Developers had never heard of many attacks prior to our demonstration.}
Based on developers' feedback and our analysis (as described in Section~\ref{method:label}), developers had not heard of \textit{3.7/7.0} attack categories prior to our demonstration. Moreover, only 10/23 developers were introduced to more than half of our demonstrated attacks.
Nevertheless, after the demonstration, developers expressed that they gained new insights into important attacks they did not consider but would be helpful for their future development. As P9 remarked regarding the software side-channel attacks used to infer keystrokes~\cite{su2024remote}, \textit{``I’ve built keyboards in VR, and I will think about the possibility of these attacks now.''} %

\newpar{Developers’ attack awareness undermines their perceptions of attack importance and practicability.}
\label{cognitive_bias}
Developers assigned lower scores to attacks that they did not know before.
To investigate how awareness influences perceptions, we categorized ratings by attack awareness and applied Mann-Whitney U-tests to assess significant differences.
Our analysis revealed a significant difference ($U = 4245.5$, $z = -3.458$, $p < 0.001$) between developers' ratings and awareness of attacks. Developers rated known attacks higher median importance score of 7.0/7.0 versus 6.0/7.0 for attacks they were unaware of before.
Similar to attack importance, for attack perceived practicability, developers gave significantly higher ratings to the attacks they were aware of ($U = 5032.5$, $z = -6.125$, $p < .001$), with a median score of 7.0/7.0 compared to 5.0/7.0 for those they were unaware of before.

\subsection{What makes attacks overall more important in XR}

After being presented with various attack types (Table~\ref{table:attacks_in_xr}) and their implications, developers rated attack importance with a median score of 6.0/7.0. We analyze their reasoning and identify the key factors as follows:

\label{result_attack_importance}

\newpar{Immersive experiences are considered highly personal.}
P2, P8-10, P12, P14, and P16-22 noted that attacks like social attacks are particularly important because \uline{XR experiences are highly personal}. P12 stated: \textit{``It's a completely different feeling when people are in your personal space, you're using voice and you can hear them, you can talk to them, you can see their hands. It's very different than a social media app that you can ignore on a screen ... I think the damage can definitely be a lot more than other media, like I had a kid that came up to me and just sneezed in my face, and I felt so disgusted, I know, like it's virtual, and it was like: dude. Get away from me.''} 
P12, P14, and P22 emphasized that social attacks, from these previous negative experiences, can be highly disruptive and should be mitigated in both their application and other multiplayer XR applications. As P22 noted: \textit{``Especially social XR platforms like VRChat, at least consider it as part of the game’s design.''}

\newpar{XR functionalities may negatively impact the large user population.}
P6, P10, P19, P22 and P23 identified attacks as important due to potential \uline{large-scale sensitive information leakage} from the community's perspective.
As P6 stated about the shoulder-surfing attack using XR devices: \textit{``This sounds like a very big issue. 
Because you know AR glasses are coming soon right? By a bunch of companies, or you know what, a lot of them have already been around. I think mass adoption is going to be coming within the next couple of years.''} %
Similarly, without hesitation, P4, P6, P7, P9-11, P16, and P21 believed social attacks are the most important attacks in XR because they affect a large number of users. To avoid similar threats happening in their applications, developers stated that they would actively consider more of these attacks during implementation. For example, P20 mentioned 
 their efforts to reduce the risk of social, physiological, and perceptual attacks:\textit{``That's why we're always trying to consider that in our design. And we're gonna do a lot like testings internally before we bring it to test down users.''} Moreover, P8, P15, P16 and P18 mentioned that developers are worried that negative experiences caused by these attacks could drive users away.

\newpar{XR applications are beginning to carry out higher-risk activities.}
P3, P4, P7, P8, P13, P14, P16, and P21-23 indicated that attacks like side-channel attacks could be used to infer sensitive user information during \uline{higher-risk activities} in XR, such as password entry, private behaviors, conversations, or other personal details, making these attacks especially important to consider in XR development. P16 commented: \textit{``Those can track a lot more physical data, and also your health data, and even the like. Your motions, signatures, or even capture, if you have certain injuries.''}
Similarly, P13 mentioned that this threat will become more severe in the near future as higher-risk operations, such as banking, are introduced in XR. Developers and the XR community should prioritize their focus on protecting data used in these higher-risk applications.

\newpar{The immersive nature of XR makes mitigation more challenging.}
P6, P16, P19, and P22 believe that attacks are more important in XR as they are harder to mitigate for developers, which aligns with~\cite{castro2022_arvr_moderation}. P6 used social attacks as an example: \textit{``You probably aren't able to do real-time voice detection.''} 
Similarly, P16 commented: \textit{``I think it is very important ... I also think it's difficult to moderate, most things like Rec Room, they're unmoderated experiences.''}

\subsection{What makes certain attacks less important in XR}
\label{attack_less_important}
While developers generally perceive attacks as important in XR, they consider some attacks less important (score $<$ 4). 
We analyzed their reasoning
and the key factors as follows:

\newpar{Users are capable of performing mitigation.}
As highlighted by P3, P7-9, and P14, physiology attacks can be perceived as less important since users can simply remove headsets or close their eyes for mitigation instead of implementing mitigation during the development process. For example, P8 mentioned:\textit{``
 I think it's too easy to just take off the headset. It's not like I'm holding you hostage or anything.''} Similarly, P7 mentioned that closing eyes can also prevent these attacks.
However, this shifts the burden of mitigation onto users, which contradicts recommendations for preventing physiology attacks~\cite{cayir2023augmenting, porcino2022guideline}.

\newpar{Certain attacks are perceived as applicable only within specific XR environments.}
P4, P6, P7, P10, P16, P17, P20, and P21 mentioned that they consider certain attacks less important when they believe those attacks are only \uline{effective under certain circumstances}. For example, P4 commented on physiology attacks: \textit{``I think the damage is more or less minimal because it's impacting an individual user ... in most cases it's more so harmless and just a Disney effect.''}

\subsection{What makes attacks overall more practical in XR}
Similar to the process of evaluating attack importance, we asked developers how practical the attacks were and identified a median Likert score of 6.0/7.0. We
analyzed their reasoning and identified the key factors as
follows:

\newpar{XR provides stealthiness and flexibility through hardware and immersive technology.}
P6, P8-10, P13, P14, P16, P19, P21 and P22 perceived attacks (e.g., shoulder-surfing attacks) as practical because of their \uline{stealthiness} to perform attacks when comparing XR to other mobile devices. P13 mentioned: \textit{``If you're using a mobile or a camera. People at least will get an awareness, right? Somebody is trying to steal something from me. Different cameras again, it's hard. But this one is like an easy option for a threat, and hard to identify, for the person who is being watched.''} %
Moreover, P6, P8-10 and P16 mentioned that social attacks are more \uline{flexible} in XR, making them more practical from a technical perspective. Based on these experiences, P16 and P19 emphasized that they want to contribute to mitigation for better user education and improved notification systems by developers and platforms.

\newpar{Encouraging user-generated content in XR may enable adversaries.}
P4, P9, P16 and P22 noted that since XR environments encourage users to create their own content (e.g., game rooms), the likelihood of physiology attacks increases. P4 noted: \textit{``I think in VRChat, you can throw a flash bomb to introduce seizure...We rely on the users to like deal with this. Platform allows these kinds of behaviors.''}
P4 was also concerned with content attack introduced by user-created content: \textit{``That's actually one of the things we tried to prevent when we're doing an event on Horizon, because the user can just jump on stage or show models or images that's not great.''} Considering this issue, P4, P16, and P22 have mentioned the need in their applications for better access control and moderation tools to mitigate malicious user-generated content.

\subsection{What makes certain attacks less practical in XR}
\label{attack_less_pratical}

While developers mostly agreed attacks were practical, some gave lower scores ($<$ 4). We analyzed their reasoning and identified the key factors as follows:

\newpar{Execution of attacks requires significant technical expertise.}
P10, P12, P15, P17, and P23 mentioned that attacks, such as perception and input attacks, are \uline{technically difficult} in XR, making them less practical. For example, P15 highlights perception attacks: \textit{``You need a lot of technical know-how to be able to manipulate boundaries like that.''} %
Similarly, P4, P7, P9, P15, and P16 noted that conducting input attacks requires technical expertise in XR or an understanding of users' states. 

\newpar{User-experience oriented attacks are perceived as less motivating.}
P6, P8, P13, P14, P16, P17, P21, and P22 mentioned that attacks (e.g., perception attacks), seem less motivating for attackers, as they primarily \uline{target user experiences} rather than direct benefits, making them impractical. As P6 commented: \textit{``You can't really gain anything from this. Aside from harming people, basically like you can't earn money from it right?''}

\subsection{Potential misconceptions on security in XR}
Based on the developers' perceptions above, we analyze and summarize the following misconceptions about XR security:

\noindent\newpar{Misconceptions on attack importance.}
From Section~\ref{attack_less_important}, we identified the following misconceptions: (1) considered that attacks that can be mitigate by users as less important in development.
While users can sometimes mitigate attacks (e.g., removing a headset during a physiology attack), developers should not shift the responsibility to users. Instead, they should implement robust defenses, especially when users may be unaware of potential threats. This view is supported by~\cite{wijayarathna2018responsible,gorski2022eight}.
(2) Misconception that 
the attacks only affect specific groups (e.g., physiology attacks on individuals with seizures) are less important. Developers should not overlook these risks. All users should be considered equally in security design, especially when user safety is at stake~\cite{wang2018inclusive}.

\noindent
\newpar{Misconceptions on attack practicability.}
From Section~\ref{attack_less_pratical}, we identified the following misconceptions:
(1) Attacks with technical complexity are perceived as less practical. A more rigorous approach to assessing attack feasibility involves using a threat model that assumes attackers possess the necessary skills and have reasonable access to relevant information~\cite{shostack2014threat}.
(2) Developers may underestimate attacks due to a lack of appreciation for attacker motivations. Specifically, they may overlook user experience–oriented attacks, assuming they are less likely. However, as XR technology is increasingly deployed in critical domains such as the military~\cite{anduril_meta_xr_2025} and healthcare~\cite{letsnurture2025arhealthcare}, these types of attacks remain both feasible and attractive to adversaries.
Moreover, all user deserves full protection, as emphasized in ethical development guidelines~\cite{gogoll2021ethics}.

\section{Developers' perceptions on current mitigation practices and supports on XR \SP}
\label{RQ2}
After examining developers’ perceptions of \SP threats in XR, we aim to understand whether they are capable of addressing these threats with appropriate mitigation strategies. %
Based on developers' responses to threats and our questions, we address our second research question by summarizing the findings that answer:
(1) What are the limitations of the current mitigation practices for XR \SP threats?
(2) What are the missing supports in current mitigation practices for XR \SP threats?
Specifically, we present insights on these questions from the perspective of both the developers and the broader XR community (i.e., other stakeholders). 
This is important because we aim to identify the limitations of current mitigation practices and address existing problems. Moreover, we seek to understand how other stakeholders in community can better support developers in ensuring \SP.

\subsection{Limitations of current practices}
\label{suboptimal}
As a first step, we identified limitations in proposing mitigations, awareness of existing mitigation practices, and perceptions of those practices provided in XR communities.

\newpar{Developers lack the ability to propose effective mitigation strategies.}
After demonstrating each attack, we asked developers to propose a mitigation strategy and evaluated their responses using the criteria described in Section~\ref{method:label}. Developers sometimes provided impractical or unclear mitigation strategies.
Additionally, developers' awareness of attacks influenced their ability to propose effective mitigation strategies. %
We conducted a Mann-Whitney U-test to compare mitigation scores between the two groups. Developers provided significantly better mitigation for attacks they were aware of, rating familiar attacks higher ($U = 3956.5$, $z = -2.48$, $p < .01$). 

\noindent\newpar{Many developers are unaware of existing policies and standards that inform best practices for privacy in XR.} 12/23 developers were unaware of the GDPR~\cite{gdpr}, and 14/23 were unaware of the CCPA~\cite{ccpa}, both of which are relevant to \SP in XR applications.
Moreover, none were aware of XR-specific standards such as Rosenberg et al.~\cite{rosenberg23the}.  %

\newpar{Developers are unaware of existing mitigation practices for XR.} %
When presented with mitigation tools for \SP in XR development, developers noted that they were unaware of some of these tools but believed these tools could be helpful in addressing \SP challenges in XR. For example, 14/23 developers reported unfamiliarity with program analysis tools (e.g., Ghidra~\cite{ghidra} and Roslyn~\cite{roslyn}) used for detecting vulnerabilities, and 14/23 were unfamiliar with existing system- or game-engine-level protection methods (e.g., Arya~\cite{lebeck2018arya}).

\newpar{Existing mitigation sacrifices utility in exchange for \SP.} 
P3-6, P12-14, and P21 expressed concerns that existing privacy mitigation practices (e.g., restricting access to camera feeds) constrain developers’ creativity and limit the potential of XR applications. P6 noted:
\textit{``They don't provide any access to the camera and information at all, It really limits what you can develop as a developer.''}

\newpar{Platforms need to improve the \SP checking process for application publication.} 
Developers highlighted that all current platforms publishing \SP policies need improvement. In our study, 14/23 participants had published applications on the Meta Quest platforms. Among them, four reported that their submissions did not undergo a detailed review of \SP, and five recommended that Meta enhance its review process by implementing more rigorous content checks and providing clearer feedback, guidelines, and documentation.
Compared to Meta, fewer developers in our study (6/23) had published on the Apple platform. However, the satisfaction rate was slightly higher: only two of these six developers believed that Apple could benefit from stricter reviews or more developer guidance, while the remaining four considered Apple’s policies to be adequate. Additionally, P11, P15, P19, and P22 observed that \SP requirements are largely absent on other distribution platforms (e.g.,  SideQuest, itch.io, and AR platforms).

\newpar{Disclosure and feedback on \SP problems is not sufficient.}
P1, P4-10, P13-16, P19, and P22 emphasized the importance of timely disclosure of \SP issues to ensure effective mitigation, but noted that such disclosure is largely lacking at present. To address this problem, developers called for the establishment of more testing and disclosure channels—both formal (e.g., beta testing, bug bounty programs) and informal (e.g., app reviews, user comments)—to help identify implementation and design flaws related to \SP issues. For example, P15 noted that collaboration and disclosure among developers was invaluable in his previous XR development projects at larger companies. \textit{``Have developers that are all working on like VR or XR in general, when you publish it, and they all test it, and tell you what's good, what's not, and then you fix that... We need dedicated developer-focused forums, similar to Reddit, to gather feedback from experienced developers. This feedback can help you refine your application and enhance its privacy compliance.''}
Moreover, P13 called for additional tests from users for \SP issues in their applications and help uncover previously unknown problems.\textit{``I would like users to be open about the issues they are facing. However, I haven't gotten any feedback provided by my end users regarding the protection mechanism used in the application.}''

\newpar{Regulations and best practices for XR are impractical or outdated.}
 P1-11, P13-16 and P21, P22 expressed frustration with existing guidelines and documentation in XR by describing them as consistently difficult to follow. They characterized these resources as overly lengthy, frequently updated, and occasionally inaccurate which is largely attributed to the current stage of the XR industry. P16 also noted that this issue stems from the suboptimal practices of industry-leading platforms: \textit{``Meta is the biggest vendor out there, and they have so much old documentation mixed with new documentation. Oftentimes it is impossible to know what is current, and sometimes even the current stuff is not well documented.''}

\subsection{Absence of supports from the XR community}
\label{missingsupports}
In addition to the limitations of existing solutions, developers' suggestions for mitigating threats and other suboptimal practices in XR (e.g., outdated documentations) revealed the lack of support within the XR industry. We categorized these needs into three groups demonstrated below:

\subsubsection{Strategic Support}
\label{strategic}
From the developers' response, we identify a need for strategic support (i.e., the provision of guidance or resources) as follows.

\newpar{Better support for developers to address economic and utility concerns in \SP-aware XR development.} %
 P2, P6-10, P12, P13, P15-19, and P21 highlighted that the tension between utility and security is fundamentally rooted in the economic constraints of XR development. P7 remarked:"\textit{Security is definitely something we want to consider sooner rather than later, but it's essential to have more cost-effective solutions.}" Moreover, as previously discussed, developers called for strategies different from existing ones (e.g., blocking the camera feed) that do not compromise utility. P2 and P11 also emphasized the need for strategic support to balance the economic needs of smaller companies with \SP in XR:\textit{``[Companies/Policymakers] should reward developers that demonstrate greater responsibility toward the technology, to encourage them to build a compliant application.''} P5, P7-9, P16, and P22 also emphasized the need for community to provide external \SP experts, which developers can not afford.

\newpar{Better guidance for best \SP practices in XR.} 
After demonstrating the threats, P1-3, P5-13, P15, P16, and P22 emphasized the need for guidance and professional resources to address emerging attack surfaces in XR. %
For example, P5 and P16 highlighted the responsibility of industry leaders (e.g., Meta, Apple) to provide accessible resources such as checklists, templates, tutorials, and consulting services to support developers, particularly junior developers, in \SP practices such as privacy policy drafting. %
Moreover, developers found that general \SP policies (such as GDPR and CCPA) were often tedious, costly to implement. For example, P8 mentioned:\textit{``The biggest problem I have with the GDPR is that it's so annoying that I would beg to have it go away.''} Moreover, P20 mentioned that many requirements might not be suitable for XR:\textit{``I don't made any experience that's telling the people ahead that we're tracking your motion data because that's kind of ruins the magic of the [XR] experience.''}  Based on this, P2, P4, P6, P8, P11, and P18-22 called for more XR-specific regulations to address the unique use cases (e.g., notification, immersive social experiences) in XR applications.

\noindent\newpar{Better consideration for underrepresented groups}. All female developers (P1, P11, P18, and P19), raised concerns about insufficient support in XR for underrepresented groups. P11 pointed out that physiological attacks disproportionately affect women, since many XR testing focus primarily on male participants, and called on developers to address this imbalance. This mirrors our finding here that only female participants flagged these issues. P19 described having experienced social attacks: \textit{``I got attacked a couple of times, especially when your avatar is a woman or a minor''} and urged stronger anti‑discrimination measures and requirements in social XR experiences.
Besides the worry of female developers, many developers (P6-13, P16, P17, P19, P21, and P23) emphasized that current XR platforms lack adequate protections for children, underscoring the urgent need for age‑appropriate security safeguards in the XR community.

\subsubsection{Technical Support}
\label{technical}
Developers also called for stronger technical support as follows:

\newpar{Moderation tools:} P2, P5, P8, P9, P11, P13, P14, P16, P17 and P19-22 requested moderation tools (e.g., vision-based) for content and user behavior management in XR applications. P13 specifically highlighted the need for integrated moderation tools to block malicious content.

\newpar{I/O processing tools:} P2, P6, P7, P8, P10, P11, P14, P16, and P19 emphasized the need for input/output processing tools.
For instance, P6 suggested differential privacy-based automatic blurring as an output processing method for mitigating shoulder surfing attacks, while P8 highlighted that input checking, such as packet inspection can help prevent input attacks.

\newpar{User education and notification systems:}
P1, P4, P6-10, P13, P16, P17, and P19-22 emphasized the need for improved user education and notification systems as a technique to mitigate content, social, shoulder-surfing, input, and physiology attacks. For example, P16 stressed the importance of implementing an interaction design to notify users of content attacks.

\subsubsection{Communication Support}
\label{comunication}
Developers also identified gaps in communication support as follows:

\newpar{Communication for developers to understand XR \SP knowledges.}
P1, P2, P4, P5, P7, P11-13, P16, and P18-22 stressed the importance of providing more threats knowledge with clear guidance on \SP in XR development. P2 highlighted: \textit{``When I see, like XR related posts it's always about. Oh, what's the new tech like AI and stuff like that? But it's never about like security. So I think, having the community participate and educating everyone about it, and then giving suggestions.''}
Moreover, all developers hope researchers could share findings promptly and educate the broader community through accessible channels (e.g., online tutorials and developer summits). 
P1, P2, P8, P15, and P19 also mentioned the need for more communication channels dedicated to \SP. The absence of such channels may also contribute to the existing communication problems (e.g., disclosure on \SP issues).

\newpar{Communication for technical transparency}
(P2, P4, P5, P6, P7, P8, P9, P11, P12, P13, P14, P16 and P19).
Developers expressed concerns about the lack of communication on technical transparency in the XR ecosystem, particularly regarding the black-box nature of third-party APIs (e.g., Avatar SDK, XR Tracker), which they feared could introduce \SP risks. 
Participants emphasized the need for clearer communication, reliability, and openness from third-party providers.

\subsection{Analysis: Misconceptions on the supports for \SP in XR}

Although most perceptions of unmet needs are valid, our analysis also identifies the following misconceptions.

\noindent
\newpar{Misconceptions on policies and standards.} Developers mistakenly believe that there are no XR-specific \SP policies or standards. While we agree that developing more XR-specific \SP policies is beneficial, as discussed in Section~\ref{suboptimal}, there are already some initial efforts in this area. Addressing developers’ lack of awareness of existing policies is also a crucial part of the problem~\cite{gdpr,ccpa,rosenberg23the}.

\noindent
\newpar{Misconceptions on technical support.} Developers also mistakenly believe that there are no usable tools for mitigating the proposed threats (e.g., content attacks). While we agree that additional tools are needed to address XR \SP threats, there have already been initial efforts and tools made available to developers~\cite{nair2022going, ghidra, roslyn}. However, a lack of awareness among developers about these existing mitigation tools significantly limits their effectiveness.

\noindent
\newpar{Misconceptions on communication.} Besides, developers hold the misconception that there are no usable communication channels for XR \SP. While we agree that more communication channels and feedback mechanisms would be beneficial, there are existing platforms—such as forums~\cite{meta_community_developer_forum, nvidia_xr_spatial_forum, immersiveunity_xrcommunity} and conferences~\cite{awe2025, ieeevr2025}—that already serve this purpose. Developers should proactively engage in \SP discussions through these channels, participate in bug bounty programs, and follow up with users to close the feedback loop.

\section{Discussion}
\label{discussion}

In this work, we investigate how developers perceive and reason about real-world XR \SP threats. By examining their perceptions of the sensitivity, realism, importance, and practicality of threats, we uncover gaps in development (see Section~\ref{result_privacy} and~\ref{result_security}) and highlight key factors (see Section~\ref{RQ2}). Moreover, our comprehensive study reveals two common misconceptions: (1) awareness gaps regarding XR threats, and (2) diffusion of responsibility when discussing mitigation practices for XR threats. In this section, we further examine the consequences of these misconceptions, analyze their underlying causes, and propose corresponding solutions. We also discussed the limitation and future directions in Appendix~\ref{limitation}.

\subsection{Awareness gaps due to the rapidly evolving nature of XR interaction technologies.}
In Sections~\ref{awareness:privacy},~\ref{awareness:security}, and~\ref {suboptimal}, we identified significant gaps in developers’ awareness of emerging XR threats, existing mitigation tools, and support resources (e.g., reward programs~\cite{meta_metaverse_responsibly}, guidance~\cite{XRSI_2023}).
These awareness gaps lead to \uline{normalcy bias}—developers tend to assume the threats as safe because they are unaware of potential risks (e.g., perception attacks are widely underestimated, especially in high‑stakes contexts like military or medical XR)~\cite{chattopadhyay2020tale,securitymagazine2021normalcybias}.
This underestimation reduces developers’ motivation and capacity to implement mitigation, increasing the likelihood of suboptimal practices such as insecure interaction designs (e.g., poorly considered voice control in immersive contexts) and missing safeguards (e.g., unmoderated social XR). As noted earlier, developers identify these factors as amplifying XR threats.

We believe that this awareness gap exists due to the emerging nature of XR interaction technologies, where threats are closely tied to new interaction designs, especially the new modalities of data (e.g., motion, voice) 
and the heavy reliance on them to fully enable basic interactions in XR, rendering an inherent and difficult privacy-utility tradeoff.
Knowledge about these emerging XR threats and mitigation support is difficult for developers to access because it evolves rapidly and is scattered across various sources. Unlike mature platforms such as mobile or PC, where established guidelines are readily available (e.g.,~\cite{android_security_best_practices, scarfone2008sp}), XR \SP findings are typically shared through academic publications, industry reports, or informal developer-to-developer communication, requiring developers' active engagement.

Hence, we advocate for the development of standardized communication channels for XR developers~\cite{krauss2021current,XRAState2023}, especially those integrated into tools that support the normal development workflow (e.g., IDE), as prior research in the mobile context suggests that these just-in-time reminders contextualized in the app implementation help raise developers' awareness and encourage the adoption of secure and privacy-preserving designs~\cite{li2018coconut}.
We argue that it is important to empower developers to proactively follow emerging XR trends, engage in community discussions, and utilize existing channels (e.g., forums~\cite{metaCommunityForum}, feedback systems, and app reviews) to stay informed of new threats and mitigation supports.

\subsection{Diffusion of responsibility due to the user-experience oriented nature of XR threats.}
From developers' perceptions on threats and mitigation (Section~\ref{result_privacy},\ref{result_security},\ref{RQ2}), we identified a trend of diffusion of responsibilities~\cite{darley1968bystander} (i.e., the tendency to assume that addressing XR \SP issues is someone else's responsibility). For example, developers suggested that users can mitigate physiology attacks by removing the headset, a problematic stance that disrupts the user experience and shifts both responsibility and risk onto users. Although users and platform providers also share responsibility, developers play a crucial role in collaboratively addressing these risks~\cite{naiakshina2017developers,felt2012android}.

We believe this diffusion of responsibility is especially pronounced in XR because threats (e.g., physiology attacks that induce motion sickness) span both hardware and software domains and directly impact the users' body and mind~\cite{roesner2014security,giaretta2024security}.
Unlike traditional exploits (e.g., remote code execution) that map cleanly onto traditional vulnerability taxonomies, XR threats often look like UX, hardware, or platform issues, causing a misalignment in developers' mental models for threat identification and mitigation.
Notably, although major platform providers develop built-in \SP protection, emphasizing considerations such as data minimization, transparency, control, and on-device data processing~\cite{apple2024visionproprivacy}, they are inevitably limited to relatively coarse-grained mitigations.
For instance, although Apple Vision Pro confines camera and sensor inputs (e.g., eye-tracking) to on-device processing, an application-level gaze-driven UI still reveals where users are looking through button selections—effectively leaking the eye-tracking data~\cite{smith2024privacy}.
Failing to effectively engage developers to mitigate threats represents a missed opportunity to appropriately balance \SP considerations with the nuanced demands of functionality and the complexities of social contexts.

We advocate for the development of a practical industry framework for XR that clearly assigns ownership of emerging threats to designated stakeholders and provides \SP checklists for each threat. This approach has proven effective in encouraging responsible development in other domains, such as the Web~\cite{owaspTopTen}, mobile~\cite{alkindi2021user}, and IoT~\cite{brass2018standardising}. While there have been initial efforts in this space (e.g., XRSI’s recommendations~\cite{XRSI_2023}), our findings highlight a pressing need for:
(1) more detailed practice guidelines that address real-world threat scenarios encountered in XR (e.g., social harassment~\cite{alhakamy2024extended}, motion as a side channel~\cite{su2024remote}), similar to the OWASP Top Ten~\cite{owaspTopTen} in web development; and
(2) accompanying enforcement mechanisms to motivate secure development practices (e.g., incentives or penalties from policymakers for \SP, additional testing support from users, and auditing tool development from researchers and organizations).
Such an approach is critical for fostering greater awareness among developers about their \SP responsibilities and for motivating proactive ownership of XR threats.

\section{Conclusion}

In this work, we conducted an interview study with 23 professional XR developers to investigate their perceptions and responses to threats and mitigation in XR development. This study represents an initial effort to address XR \SP issues through a developer-centered, threat-aware perspective.
Our findings reveal that developers recognize XR \SP risks as closely tied to, and often amplified by, development decisions. However, there is limited awareness of these issues among developers, compounded by a lack of strategic, technical, and communication support from key stakeholders within the XR community.
To address these issues, we analyzed developers' reasoning around threats and mitigation strategies to uncover existing problems in XR development and propose actionable, stakeholder-aware solutions. Our proposed future directions clarify responsibilities for key stakeholders and encourage stronger collaboration across the XR ecosystem. The insights from this study offer valuable directions for future research and practical advancements in XR \SP.

\section*{Acknowledgment}
We sincerely thank the reviewers for their valuable feedback on the paper.
This work is supported in part by the National Science
Foundation (NSF) Awards 2323105, 2317184. Any opinions, findings, conclusions, or recommendations expressed
in this publication are those of the author(s) and do not
necessarily reflect the views of sponsors.

\bibliographystyle{IEEEtran}
\bibliography{IEEEabrv,ndss26}

\appendix

\subsection{Limitations and Future Work}
\label{limitation}
Since our study involve demonstrations of XR risks, there is a potential for bias in developers' perceptions of \SP. However, this is unavoidable as our aim was to capture their views on broader XR \SP threats, despite possible gaps in their knowledge. It is also a common practice for research in threat-aware developer studies on other platforms~\cite{redmiles2020comprehensive, cummings2021need}.

Our work takes the first step towards understanding developers' perceptions and challenges for building \SP-friendly XR apps. There would be many valuable future directions building on top of our work, such as vetting developers' actual implementations for potential \SP issues, and development frameworks that support \SP-by-design XR developments.

\begin{table*}[h]
\fontsize{7.5}{7.5}\selectfont
\centering
\caption{Mitigation tools and best practices for XR security and privacy}\label{table:approaches_xr_security_privacy}
\begin{tabular}{clll}
    \toprule
    \textbf{No.} & \textbf{Approach} & \textbf{Examples}  & \textbf{Applicable to}\\ \midrule
    1 & User notification\&education & Privacy policy~\cite{guo2024empirical,trimananda2022ovrseen}, in-app notification~\cite{zenner2018immersive}, etc. & Content attacks, Perception attacks, Data leakage, etc.\\ \midrule
    2 & Static analysis & Ghidra~\cite{ghidra,guo2024empirical}, Roslyn~\cite{roslyn}, etc. & Software side-channel, Content attacks, Data leakage, etc.\\ \midrule
    3 & Runtime analysis & Moderation~\cite{sabri2023challenges}, packet analysis~\cite{lyu2023metavradar, trimananda2022ovrseen}, etc. & Social attacks, Physiology attacks, Data leakage, etc.\\ \midrule
    4 & OS and Game Engine enforcement & Arya~\cite{lebeck2018arya}, OS level data control~\cite{metaprivacy}, etc. & Content attacks, Input attacks, Perception attacks, etc.\\ \midrule
    5 & Legislation and policymaking & Laws~\cite{ccpa,gdpr}, best practices in XR~\cite{XRSI_2023,vrc}, etc. & Shoulder surfing attacks, Social attacks, Content attacks, etc.\\ \midrule
    6 & Input and output processing & Input or output checking~\cite{roesner2021security,lebeck2016safely}, differential privacy~\cite{nair2023truth}, etc. &Software side-channel attacks, Shoulder surfing attacks, etc. \\ 
    \bottomrule
\end{tabular}
\end{table*}

\begin{table}[h]
\fontsize{7.5}{7.5}\selectfont
\centering
\caption{Data Leakage Channels in XR we Collected From Literatures and Discussions Like~\cite{su2024remote, zhang2023s,giaretta2024security}. }\label{tab:channel}
\begin{tabular}{cp{3cm}p{4cm}}
    \toprule
    \textbf{No.} & \textbf{Leakage Channel} & \textbf{Example} \\ \midrule
    1 & App designed functions & Motion data for rendering avatar \\ \midrule
    2 & Hardware side-channels & Rendering side-channels \\ \midrule
    3 & Third-party services & Analysis APIs \\ \midrule
    4 & Insecure operations & Unsafe storage, unencrypted data \\ 
    \bottomrule
\end{tabular}
\end{table}

\subsection{Labeling Criteria and Examples}
We provide additional details to elaborate on our labeling criteria from Section~\ref{method:label}.

\newpar{Awareness of attacks:}
Participants were labeled ``aware'' if they met any of the following criteria; otherwise, ``unaware''.

\noindent (1) Proactively mentioned attacks, e.g., discussing social attacks or experiences before the demonstration of this attack.

\noindent (2) Proactively provided additional or similar examples during demonstration, e.g., mentioning user-generated light bombs triggering seizures (physiological attack).

\noindent (3) Clearly stated familiarity during or after demonstration, e.g., sharing shoulder-surfing experiences or knowing developers with mitigations for physiological/social attacks.

\newpar{Awareness of mitigation:}
Participants were labeled ``unaware'' if they met either criterion below; otherwise, ``aware'':

\noindent
(1) Explicitly stated unawareness or requested details during/after presentation, e.g., first-time hearing about mitigations like program analysis, or asking for resources.

\noindent (2) Misused tools in the matching task, e.g., suggesting Ghidra~\cite{ghidra} (static analysis) for dynamic social attacks, indicating misunderstanding.

\newpar{Quality of developer-proposed mitigation:}
We rated proposed mitigations on a three-point scale:

\noindent(1) Score 1 - Ineffective or inapplicable: e.g., no solution provided (e.g., stating they do not know), vague or high-level (e.g., \textit{``Yeah, just have good security measures.''}), or irrelevant (e.g., for input attacks: \textit{``Discomfort is like, okay, just stop using the Headset for a minute. Let the situation fix itself''}).

\noindent (2) Score 2 - On track but unclear/missing details: e.g., for content attack: \textit{``This is honestly on the developer side, like, be mindful on the way you create apps to ensure that nobody can piggyback off of what you've created, and get access, and actually show their content in your content.''} (correct direction but lacks specifics like API checks or program analysis). For input attacks: \textit{``Maybe if there was a way to disable the hand overlay, and so you can see your true hands had passed through, or something. I'm not too certain about this one this time.''} (suggests moderation but unsure/lacks details).

\noindent(3) Score 3 - Valid, clear, aligns with prior work or XR applications: e.g., for physiology attacks: \textit{``From my personal experience, it was like efficient use of data to ensure that, like on the developer side, like, you're ensuring that your data is being handled correctly, so that there's no overhead and not a lot of overload on the system already, because by itself is very intensive, especially since you're rendering every scene twice. So you have to be mindful of like, what kind of assets you use, and if they're like. compliant or not. As for the other side, if somebody else is trying to do that, like DDoS attacks and stuff like that, honestly, that's security, and like networking.''} (multiple valid solutions with examples).

\subsection{Appendix: Finalized Codebook}
\label{appendix:codes}
\begin{itemize}
\item RQ1:
\begin{itemize}
    \item Awareness of Threats
    \begin{itemize}
        \item Mentioned without prompt
        \item Provided additional examples
        \item Clearly mentioned awareness of attacks
    \end{itemize}
    \item Awareness of Mitigation
    \begin{itemize}
        \item Specifically mentioned unaware of a mitigation.
        \item Misuse mitigation tools in follow-up questions
    \end{itemize}
    \item What makes data more sensitive
    \begin{itemize}
        \item Immersive interaction design
        \item Advanced tracking sensors
        \item Economic value of user data
        \item Potential surveillance by XR platform
        \item Opaque XR infrastructure
        \item Other (e.g., data abuse)
    \end{itemize}
    \item What makes certain data less sensitive
    \begin{itemize}
        \item Significant benefits overshadow risks
        \item Insufficient knowledge of developers
        \item Don't believe data can leak sensitive information
        \item Other (e.g., also collected in other platforms)
    \end{itemize}
    \item Why data leakage channels more realistic
    \begin{itemize}
        \item Development misoperations
        \item Relied on 3rd party APIs
        \item Misused by user
    \end{itemize}
    \item Why certain data leakage channel less realistic
    \begin{itemize}
        \item Developers unfamiliar with the channel
        \item Unrelated to software development
        \item Local apps are safe
    \end{itemize}
    \item What makes attacks in XR overall more important
    \begin{itemize}
        \item Immersive experiences increase attack severity
        \item Affect the growth of the XR industry
        \item Broader negative impact in XR
        \item Affect critical operations (e.g., banking) in XR
        \item Hard-to-Mitigate characteristics in immersive social environment
        \item Other (e.g., perceivable benefit for attackers)
    \end{itemize}
    \item What makes certain attacks in XR less important
        \begin{itemize}
            \item User can mitigate
            \item Only for certain scenarios
            \item Only target certain user groups
            \item Other (e.g., less important consequence)
        \end{itemize}
    \item What makes attacks in XR overall more practical
        \begin{itemize}
            \item More Stealthy attacks enabled by XR hardware
            \item Flexibility of attacks in immersive space
            \item Already seen similar attacks happening because of implementation issues
            \item User-generated content introduces attacks
        \end{itemize}
    \item What makes certain attacks in XR less practical
        \begin{itemize}
            \item Require technical knowledge from attackers
            \item Require knowledge about the usecases/environment of the  user
            \item Less motivating to attackers
        \end{itemize}
\end{itemize}

\item RQ2:
\begin{itemize}
    \item Quality of mitigation:
        \begin{itemize}
            \item Score1: The proposed mitigation is ineffective.
            \item Score2: On the right track. Not very clear, mistaken, or missing important details.
            \item Score3: Valid, clear results that align with solutions in prior research works or adopted in applications.
        \end{itemize}
    \item Limitations XR threats mitigations
    \begin{itemize}
        \item Developer: Awareness of existing mitigation
        \item Community: \SP design sacrificed utility
        \item Community: high requirement of external resources
        \item Community: Impractical or outdated resources
    \end{itemize}
    \item Developers’ General Unmet Needs from The Entire XR Community - Communication
    \begin{itemize}
        \item More Community Communication Channels:  emerging attack channels
        \item More Community Communication Channels: XR components that
are vulnerable to attacks
        \item Attack Mitigation
        \item How to implement \SP practices in XR application 
        \item Lack of open platforms for communication
        \item Potential solutions to threats
    \end{itemize}
    \item Developers’ General Unmet Needs from The Entire XR Community - Strategy
     \begin{itemize}
        \item Strategic Support : \SP hurting utility
        \item Strategic Support : utility compromise \SP
        \item Strategic Support : standard and best practices
     \end{itemize}
    \item Developers’ General Unmet Needs from The Entire XR Community - Technical
    \begin{itemize}
        \item Better Technical Solutions : moderation
        \item Better Technical Solutions : I/O
        \item Better Technical Solutions : hardware
        \item Better Technical Solutions : authentication
        \item Better Technical Solutions : notification
        \item Better Technical Solutions : Other
    \end{itemize}
    \item Specific Unmet Needs from Each Key Stakeholder in XR Development
    \begin{itemize}
        \item More Responsible API Provider : Third-Party API as leakage channels
        \item More Responsible API Provider : Malicious Third-Party API 
        \item More Support from API Providers To Handle Updates
        \item More Transparent Industry Infrastructures from XR Platform Providers
        \item More Instructions from Policymakers
        \item More Active and Responsible Publishing Platforms
        \item More Engaged and Responsible Users
        \item More Guidance from Researchers
    \end{itemize}
\end{itemize}

\item{Other Information}
\begin{itemize}
    \item Future Considerations: Challenge and Direction
    \begin{itemize}
        \item AI
        \item Hardware
        \item Children
        \item More Complicated Attacks
        \item Bias and Ethics
        \item Discuss all potential problems now, even if they're not happening yet
    \end{itemize}
    \item Developer Proposed Mitigation
    \begin{itemize}
        \item Shoulder Surfing Mitigation
        \item Software Sidechannel Mitigation
        \item Input Mitigation
        \item Social Mitigation
        \item Content Mitigation
        \item Perception Mitigation
        \item Physiology Mitigation
    \end{itemize}
\end{itemize}
\end{itemize}

\subsection{Appendix: Pre-study Survey}
\label{prestudy}
\subsubsection{XR Development Background}
\begin{itemize}
    \item Q1: In your opinion, how critical is security within Extended Reality (XR) applications in 1-7
    \item Q2: In your opinion, how critical is privacy within Extended Reality (XR) applications in 1-7
    \item Q3: Which stakeholder in the XR community do you think should be responsible for these threats (Multiple Choice)
    \begin{itemize}
        \item Developer
        \item User
        \item Policy Maker
        \item App Store
        \item Other (Please specify)
    \end{itemize}
    \item Q4: How would you rank these privacy/security problems in XR applications? (1 is the most serious problem and 5 is the least serious problem)
    \begin{itemize}
        \item Leaking PII data (e.g., biometric data)
        \item Showing inappropriate content to users (e.g., content that induces discomfort or is for adults only)
        \item Performance issue (e.g., drop framerate/crash/lag)
        \item User misoperation (e.g., click on unwanted Ads/Drop game contents)
        \item User physical discomfort/harm (e.g., hit physical object and get hurt)
    \end{itemize}
    \item Q5: Is Your Application Single-Player or Multi-Player? (Select all applies)
    \begin{itemize}
        \item Single-player
        \item Multiple-player
    \end{itemize}
    \item Q6: What kinds of applications have you developed? (Select all applies)
    \begin{itemize}
        \item Social
        \item Education
        \item Healthcare
        \item Retail
        \item Design
        \item Military
        \item Action
        \item Role-playing
        \item Puzzle
        \item Sport
        \item Protocol
        \item Operating System
        \item Other (Please specify)
    \end{itemize}
    \item Q7: What's some interesting/unique function supported by API in your application? (Please select all applies)
    \begin{itemize}
        \item Machine learning services
        \item Third-party analyzing tools
        \item Advertisements
        \item Social-related functions
        \item Video streaming
        \item Location services
        \item Audio and transcript
        \item Haptic
        \item Other (Please specify)
    \end{itemize}
    \item Q8: Where have you published your applications? (Please select all applies)
    \begin{itemize}
        \item Steam
        \item Meta (Quest store/app lab)
        \item itch.io
        \item Apple
        \item Pico
        \item Unity asset store
        \item WebXR (Please specify how you publish the app)
        \item Other (Please specify)
    \end{itemize}
    \item Q9: Which of the following best describes the applications you developed before?
    \begin{itemize}
        \item Less than 1000 downloads
        \item 1000 - 10000 downloads
        \item 10000-100000 downloads
        \item More than 100000 downloads
    \end{itemize}
    \item Q10: How many years have you been developing XR applications?
    \begin{itemize}
        \item $<$ 1 years
        \item 1-2 years
        \item 2-5 years
        \item 5-10 years
        \item $>$ 10 years
    \end{itemize}
    \item Q11: How many XR applications have you developed?
    \begin{itemize}
        \item (A slider with value from 0-40)
    \end{itemize}
\end{itemize}

\subsubsection{Demographic Questions}
\begin{itemize}
    \item Q12: Which of the following best describes you? 
    \begin{itemize}
        \item 18-25 years old
        \item 25-35 years old
        \item 35-45 years old
        \item 45-55 years old
        \item $>$ 55 years old
    \end{itemize}
    \item Q13: Do you speak English?
    \begin{itemize}
        \item Yes
        \item No
    \end{itemize}
    \item Q14: Level of Education
    \begin{itemize}
        \item Less than a high school diploma
        \item High school diploma
        \item College's/Bachelor's Degree
        \item Associate's Degree
        \item Master's Degree
        \item Doctoral Degree
    \end{itemize}
    \item Q15: Gender Identity: How do you describe yourself
    \begin{itemize}
        \item Cis-Male
        \item Cis-Female
        \item Trans-Male/Trans-Man
        \item Trans-Female/Trans-Women
        \item Gender queer
        \item Other (Please specify)
        \item Prefer not to say
    \end{itemize}
\end{itemize}

\subsection{Appendix: Interview Procedures and Scripts}
\label{interview}
The interview was divided into three main parts, followed by a post-study feedback segment, lasting approximately 90 minutes. The procedure was structured as follows:

\subsubsection{Part 1: Background Questions:}
\label{interview:background}
Participants were asked to describe their background in XR development, including any training they received related to privacy and security. Questions focused on participants' recent XR applications and the privacy and security challenges they encountered.

\begin{itemize}
    \item[\textbf{Q1.1}] Since we are going to talk about the apps that you have developed, do you mind first briefly introducing your background in developing XR applications?
    \begin{itemize}
        \item[-] What’s the training you have on developing XR applications?
        \item[-] Have you received training for privacy in XR applications development?
        \item[-] Did you have any questions about privacy when you are developing applications? 
        
        \hangindent=2em [If yes] Who did you consult when you have questions about privacy?
    \end{itemize}
    \item[\textbf{Q1.2}] Can you tell us more about the recent XR applications you developed?
    \begin{itemize}
    \item[-] What were these apps built for? Can you tell us more about the application?
    \item[-] We see that you mentioned your application used [interesting/unique function from prestudy survey], can you tell us more about its usage and purpose?
    \item[-] Were the apps developed by a team or just by yourself?

    \hangindent=2em [If it’s teamwork]
        \begin{itemize}
            \item[-]  What is your role in the development process?
            \item[-] Who is in charge of the application design? Will they also handle the security and privacy aspects of the app?
            \item[-] What’s the process of creating a privacy policy for your app and deciding what to include in the privacy policy? What are the challenges associated with creating the privacy policies?
        \end{itemize}
    \hangindent=2em [If by themselves] How do you usually create a privacy policy for your app and decide what to include in the privacy policy? Have you encountered any problems when creating the privacy policy?

\end{itemize}

    \item[\textbf{Q1.3}] Can you tell us about the security design of your applications?
    
    \hangindent=2em \textit{[Content presented on the shared screen] ``Security in XR, similar to security in other domains, involves safeguarding the integrity, confidentiality, and availability of applications. Security designs are defenses or mitigation against threats which can that may cause incidents of unexpected behavior to harm user experience. This threat can be unique to XR or general threats.''}
    
    \begin{itemize}
        \item[-] Have your applications considered any design for user security specific to XR?
        
        \hangindent=2em [If yes] 
        \begin{itemize}
            \item[-] What are these designs, and what aspect are they protecting?

            \item[-] What are some challenges you encountered when building the defense?
            
            \item[-] What are the limitations or problems that you can think of in the security design?
            
            \hangindent=2em [If no limitation] Why do you think the current design is enough?
            
            \hangindent=2em [Else] What additional protections do you think are needed for your application or other applications?
        \end{itemize}
        
        \hangindent=2em [If no] What do you think your application should consider for security protections?
    \end{itemize}

    \item[\textbf{Q1.4}] Can you tell us about the private data collection in your applications?

    \begin{itemize}
        \item[-] Please list the personal data collected by your applications and what function they have been used for.
        
        \hangindent=2em [If yes] Which of these data are more sensitive than others? Why?
        
        \hangindent=2em [If no] Can you tell us more about what other data you have collected in your applications?
        
        \hangindent=2em [If no personal data were collected, skip to the next question.]

        \item[-] What defense mechanisms has your application provided to avoid the leakage of these data? 
        
        \hangindent=2em [If there’s any protection] 
        \begin{itemize}
            \item[-] How did you design the defense mechanisms? Any challenges in designing this mechanism?
            \item[-] What are the limitations or problems that you can think of in the defense?

            \hangindent=2em [If no limitation] Why do you think the current design is enough?
        
            \hangindent=2em [Else] What additional protections do you think are needed for your application or other applications?
        \end{itemize}
        
        \hangindent=2em [If no protection] What do you think your application should consider for user data privacy protections?

        \item[-] How are users kept informed of the measures in place for preserving privacy? 
    \end{itemize}

    \item[\textbf{Q1.5}] What do you think about the security and privacy requirements from app stores' publishing process?
    
    \begin{itemize}
        \item[-] Are there any difficulties in satisfying app store rules like VRC rules? 

        \hangindent=2em \textit{[Content presented on the shared screen] Virtual Reality Checks (VRC). These are a set of standards and guidelines that developers must adhere to when submitting their Virtual Reality (VR) applications for publication on Meta's platforms. These include but are not limited to performance requirements and some security requirements.}

        \item[-] What challenges do you encounter when drafting privacy policies and ensuring compliance with relevant regulations? Please describe your experiences and any particular areas you find difficult.
    
        \item[-] Please describe your thoughts on the current app store's review process for applications. Do you believe it is overly cautious, or do you think there are additional aspects they should be checking? Explain your reasoning.
    \end{itemize}
    
    \item[\textbf{Q1.6}] How do you think about security and privacy in the current XR industry? Why?
\end{itemize}

\subsubsection{Part 2: Threats During XR Development}
\label{interview:Demo}
This subsection explores privacy and security threats in XR development, focusing on sensitive data, current policies, and real-world examples of potential risks.

\begin{itemize}

    \item[\textbf{Q2.1}] Please describe any privacy threats, such as data leakage, you are aware of in XR applications. [Pre-demo questions on developers' awareness]
    
    \begin{itemize}
        \item[-] What kinds of data collected from users in what way should be considered sensitive in XR? Can you give me some examples?
        \item[-] Why are these data sensitive in XR? 
        \item[-] What private information can be inferred from these data?
    \end{itemize}
\end{itemize}

\begin{center}
\textbf{\textit{[Here, demo Sensitive Data Collection in XR and collect interviewee's feedback via Qualtrics (Q:``Which data listed here should be considered sensitive? Please provide a rating from 1-7 where 1 is not sensitive at all and 7 is very sensitive with a short reasoning on the score.'')). See Table \ref{table:unique_sensitive_data_xr} -- Qualtrics is only being used for note-taking purposes here.]}}

\end{center}

\begin{center}
\textbf{\textit{[Here, demo Data Leakage Channels in XR and collect interviewee's feedback via Qualtrics. (Q: ``Which Data Leakage Channels is considered more realistic and important to you and your apps? Please provide a rating from 1-7, where 1 is not realistic at all and 7 is very realistic, and a short reasoning for your ratings. See Table \ref{tab:channel} -- Qualtrics is only being used for note-taking purposes here.]}}
\end{center}

\noindent \textit{After the above demo sessions:}
\begin{itemize}
    \item[-] [Confirm with participants on all showing data] Is your application collecting any of this information and for what purposes?
    \item[-] What protections have your application designed to safeguard the information it collects? Please describe these security measures in detail.
    \item[-] Have you listed all these data you used in the privacy policy and do you think they should be mentioned in the privacy policy? Why?
    \item[-] Considering the data and channels discussed, what specific improvements would you implement in your applications to enhance data privacy?
\end{itemize}

\begin{itemize}

\item[\textbf{Q2.2}] As a developer, how would you evaluate the sufficiency of the current policies governing extended reality (XR)? Are there specific areas where they fall short or excel?

\begin{itemize}
    \item[-] What actions can be taken to promote better privacy compliance within the XR community? Please provide your suggestions and rationale.
    \item[-] From a legislative standpoint, what improvements would you suggest to better support the development and management of XR technologies?
    \item[-] How can policies be adjusted or created to better assist developers in the XR field? What types of support are most needed?
\end{itemize}

\item[\textbf{Q2.3}] \textit{[Content presented in the share screen] Not strictly following SDK specifications (API misuse) can lead to security issues like:
\begin{itemize}
    \item[1.] Significant performance issues/crash (reduce immersive experience)
    \item[2.] Unexpected application behaviors
    \item[3.] Allow attacks (e.g., remote code execution)
\end{itemize}
}

\begin{itemize}
    \item[-] Do you usually check specifications for potential security issues before using the corresponding API?
    \item[-] Have you encountered any difficulties when reading/understanding the specifications (e.g., reading OpenXR/MRTK/Oculus specification)? If yes, how do you usually resolve it?
    \item[-] Any suggestions for the community to avoid these issues?
\end{itemize}

\item[\textbf{Q2.4}] \textit{[Content presented in the share screen]
\begin{itemize}
    \item[1.] Security attacks include any forms of digital manipulation and interference.  Aimed at compromising user security, privacy, and interaction integrity in digital environments.
    \item[2.] Privacy attacks infer or extract sensitive information from users or their surroundings.
    \item[3.] Note: The attacks may be not limited to active attack but also the malicious contents added accidentally.
\end{itemize}
}

Please describe any potential attacks that can affect security or privacy that you are aware of in XR applications (In the apps you developed/Played before). 

\end{itemize}

\begin{center}
\textbf{\textit{[Here, we demo the interviewee seven privacy and security attacks in XR and collect their feedback. See Table~\ref{table:attacks_in_xr}.]}}
\end{center}

\noindent \textit{For each attack, we collect feedback using 1-7 Likert-scale (1 = not at all, 2 = very unlikely, 3 = unlikely, 4 = neutral, 5 = somewhat likely, 6 = likely, 7 = very likely) questions and an open-ended question:}

\begin{itemize}
    \item[-] Do you think this attack is practical for XR applications (Rating 1-7)? Explain why or why not?
    \item[-] Do you think this attack is important for XR applications (Rating 1-7)? Explain why or why not?
    \item[-] What mitigation strategies do you recommend for this type of attack? Please explain why you consider these strategies effective.
\end{itemize}

\begin{itemize}
    \item[\textbf{Q2.5}] Could you share experiences from you or other developers who have taken these threats or problems into account when developing XR applications?  
    \begin{itemize}
        \item[-] Is this a speculation, or have you observed them doing this? 
        \item[-] [If yes], do you heard of of think of any challenges during this process?
        [showing the seven attacks in the shared screen]
    \end{itemize}

    \item[\textbf{Q2.6}] Other than these threats, have you noticed or thought of any other problems when developing XR applications?
    \hangindent=2em [If yes] What are the problems and their potential effects?

    \item[\textbf{Q2.7}] [Customization question] We noticed that your application uses machine learning and haptic APIs/services. Are you using third-party APIs for these services?

    \begin{itemize}
        \item[-] Could you discuss any potential privacy or security threats these third-party APIs might introduce to your application?

    \end{itemize}

\end{itemize}

\subsubsection{Part 3: Best Practices and Mitigations}
\label{interview: miti}

In this subsection, we demonstrate best practices and mitigation strategies for XR security and privacy, assessing their effectiveness and stakeholder responsibilities.

\begin{center}
\textbf{\textit{[Here, we demo some privacy and security related best practices and/or mitigations strategies. See Table \ref{table:approaches_xr_security_privacy}]}}
\end{center}

\begin{itemize}

\item[\textbf{Q3.1}] [Ask while demoing] Do you recognize any of these mitigation strategies or best practices before? If you have any questions, we are willing to explain them in more detail.

\item[\textbf{Q3.2}] [Ask while demoing] Can you discuss how effective you believe these mitigation approaches would be in addressing the problem? Please explain your reasoning and any potential impacts.
\begin{itemize}
    \item[-] How do you assess the adequacy of the current tools available to address these threats? Please describe in detail whether these tools meet the needs, and where gaps might exist.
\end{itemize}

\item[\textbf{Q3.3}] In the XR community, various stakeholders such as developers, platform providers, regulatory bodies, and users play different roles. Which of these stakeholders do you believe should be responsible for addressing security and privacy threats? You may identify multiple stakeholders.
\begin{itemize}
    \item[-] Can you match these stakeholders with their responsibilities? Or add more stakeholders to the list?
    \item[-] Can you explain your choices and discuss the specific responsibilities you believe each should have in managing these threats?
\end{itemize}

\begin{center}
\textbf{\textit{[Here, direct participants to Qualtrics to match stakeholder-responsibility relationships. -- -- Qualtrics is only being used for note-taking purposes here.}}
\end{center}

\noindent [Qualtrics Question] Please enter the stakeholder numbers to the field of each responsibilities (Can be multiple stakeholders in one text box)

\noindent \textit{The stakeholders include: }

\begin{enumerate}
    \item \textit{App Developer}
    \item \textit{App store}
    \item \textit{XR System/HMD Developer}
    \item \textit{User}
    \item \textit{Policymakers}
    \item \textit{Other (specify in the text box)}
\end{enumerate}

\noindent \textit{The responsibilities (with corresponding text boxes to fill in Qualtrics) include:}

\begin{enumerate}
    \item \textit{Understand the potential security threats}
    \item \textit{Understand the potential privacy threats}
    \item \textit{Limit the sensitive data sent to malicious parties}
    \item \textit{Avoid malicious content added to the application and protect user interaction integrity}
    \item \textit{Inform the user of potential threats in the application}
    \item \textit{Ensure compliance of the application}
    \item \textit{Avoid vulnerabilities in the application}
    \item \textit{Other (specify in the text box)}
\end{enumerate}

\end{itemize}

\subsubsection{Part 4: Post-Study Feedback}
\label{interview: feedback}

In this subsection, participants provide feedback on their impressions of XR security and privacy and share insights gained from the study.

\begin{itemize}
\item[\textbf{Q4.1}] Other than the discussed examples, what do you foresee as the future challenges or directions for developing safe XR applications?
\begin{itemize}
    \item[-] Why do you believe these could become future trends?
    \item[-] Do you believe other developers are aware of these future challenges, and/or has proper research been conducted to equip them with the necessary knowledge to tackle them?
\end{itemize}
\end{itemize}

\begin{itemize}
   \item[\textbf{Q4.2}] What's your overall opinion on security and privacy in XR after attending our study?
\begin{itemize}
    \item[-] Which part of our demo surprised you the most?
    \item[-] How has this study influenced your views on XR security and privacy, if at all? Please share any insights you have gained and discuss how they might be helpful in your context.
    \item[-] What information from this study do you think other developers need to know, or what would you suggest to your colleagues?
\end{itemize}
\end{itemize}

\newcommand{\developerCnt}{16\xspace}

\end{document}